\documentclass{article}
\usepackage[accepted]{mlsys2025}

\makeatletter
\renewcommand{\Notice@String}{Preprint.}
\makeatother

\usepackage{amsmath}
\usepackage{amssymb}
\usepackage{bm}
\usepackage{amsthm}
\theoremstyle{plain}

\newtheorem{proposition}{Proposition}
\theoremstyle{definition}

\usepackage{xurl}
\usepackage{booktabs}
\usepackage{graphicx}
\usepackage{listings}
\usepackage{float}
\usepackage{multirow}
\usepackage{tabularx}
\usepackage{pifont}
\usepackage{microtype}
\usepackage{tcolorbox}
\usepackage{enumitem}
\usepackage{tikz}
\usetikzlibrary{arrows.meta,positioning}

\graphicspath{{figures/}}

\providecommand{\Description}[1]{}

\newcommand{\revzk}[1]{{#1}}
\newcommand{\method}{\textsc{ReplayAudit}}
\newcommand{\E}{\mathbb{E}}
\newcommand{\bind}{\mathrm{bind}}
\newcommand{\commit}{\mathrm{Com}}
\newcommand{\link}{\mathrm{link}}
\newcommand{\sigid}{\sigma_{\mathrm{id}}}
\newcommand{\Pdetect}{P_{\mathrm{detect}}}

\mlsystitlerunning{\method{}: Auditing Outsourced LLM Fine-Tuning}

\begin{document}

\twocolumn[
\mlsystitle{\revzk{\method{}: Auditing Outsourced LLM Fine-Tuning\\via Sealed Records and Selective Replay}}

\begin{mlsysauthorlist}
\mlsysauthor{Yihan Xia}{szu}
\mlsysauthor{Wenxin Xu}{szu}
\mlsysauthor{Taotao Wang}{szu}
\end{mlsysauthorlist}

\mlsysaffiliation{szu}{College of Electronics and Information Engineering, Shenzhen University, Shenzhen, China}
\mlsyscorrespondingauthor{Taotao Wang}{ttwang@szu.edu.cn}

\mlsyskeywords{Outsourced Training Verification, Telemetry-Guided Auditing, Spot-Check Auditing, Deterministic Replay, LoRA, ML Systems Security}

\vskip 0.3in

\begin{abstract}
\revzk{Auditing outsourced LLM fine-tuning requires checking an untrusted provider's training while controlling verification cost. Uniform sampling ignores training-time signals, while checking a suspicious record can require replaying several preceding steps. We present \method{}, a telemetry-guided auditing system that seals training records before challenges and independently replays sampled intervals from sealed checkpoints. Loss changes and parameter-update norms guide stratified random sampling; replay determines whether the selected computation agrees with its records. We relate sampling probabilities to replay work and evaluate the system on real LoRA training. A 7B BF16 prototype rejects a false update-norm report during a random audit; targeted checks also reject false loss reports and skipped updates. We calibrate sampling budgets offline to at least a 90\% probability of directly selecting an injected step. For two learning-rate deviation settings, executing these budgets on the same honest 1.5B trajectory reduces median complete-audit wall time by 49\% and 67\% relative to uniform sampling. When skipped updates evade the high-anomaly stratum, the corresponding audit instead costs 38\% more. These results show how deviation capture and actual replay work jointly determine the cost of telemetry-guided auditing.}
\end{abstract}
]

\printAffiliationsAndNotice{}

\section{Introduction}
\label{sec:intro}

\revzk{Outsourcing large language model (LLM) fine-tuning lets a client adapt a model without operating its own GPU infrastructure. Parameter-efficient methods such as LoRA~\cite{hu2022lora} reduce the trainable state, but do not resolve the client's verification problem: an untrusted provider may skip updates, alter hyperparameters, or submit fabricated training records. A delivered model and a plausible loss curve do not establish that the agreed training procedure was followed. The client therefore needs an audit that checks computation while avoiding the cost of independently repeating the entire run.}

\begin{figure*}[t]
\centering
\Description{ReplayAudit connects sealed training telemetry and checkpoint records to post-seal stratified sampling and independent interval replay. An auxiliary integer checker is a separate component path.}
\includegraphics[width=\textwidth]{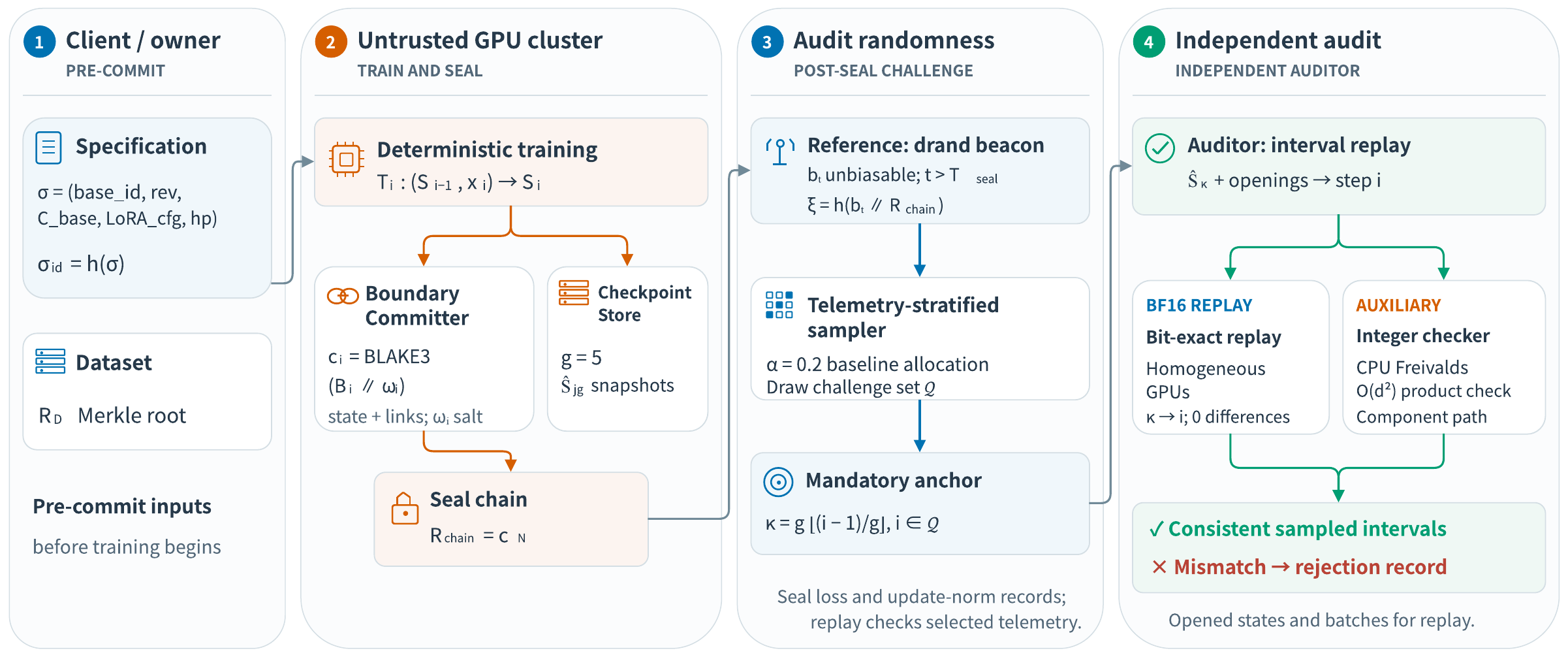}
\caption{\revzk{\method{} connects training-time recording to post-seal audit decisions. Telemetry guides target selection; checkpoint-anchored replay checks the selected computation and records. The public beacon and dataset root belong to the reference protocol; the BF16 prototype uses local post-seal randomness and batch hashes. The integer checker is an auxiliary component path.}}
\label{fig:framework}
\end{figure*}

\revzk{Existing systems provide several ways to obtain such evidence. Zero-knowledge training proofs, including VeriLoRA~\cite{verilora} and VFT~\cite{vft}, verify encoded training relations while protecting private witnesses, at the cost of proof generation. Commit-then-sample systems such as VeriML~\cite{veriml} reduce the computation selected for proof. Proof-of-Learning~\cite{pol} and AFTUNE~\cite{aftune} use selective recomputation, with different numerical and trust assumptions. These approaches establish selective verification as a useful design principle. Our question is how to use training telemetry to select effective checks, and when fewer samples translate into lower audit cost.}

\revzk{Training already produces signals such as loss changes and parameter-update norms. These signals can help direct attention toward unusual steps, but using them for adversarial auditing introduces two coupled difficulties. First, the provider controls the records: an anomalous value is not itself a violation, and a plausible value may be fabricated. Sealing prevents later edits but does not make submitted telemetry truthful. The auditor must independently check the selected computation against the sealed state. Second, a selected step can require replaying several preceding steps to reconstruct its inputs. A smaller sample set need not be proportionally cheaper, especially when replay intervals overlap or additional predecessor checks are necessary. Selection quality and replay work must be assessed jointly.}

\revzk{We introduce \method{}, a telemetry-guided auditing system for outsourced LLM fine-tuning. Our design uses provider-reported telemetry to allocate checks, while independent replay determines whether the selected records agree with the sealed computation. The provider commits step records and periodic checkpoints, then seals the trajectory before audit randomness is revealed. The auditor ranks the sealed telemetry and combines random sampling within a high-anomaly stratum with sampling from all remaining records. For each selected target, it restores a sealed checkpoint and replays the required interval, comparing states and telemetry with their recorded values. Figure~\ref{fig:framework} illustrates how these stages connect training-time recording to post-training verification.}

\revzk{To characterize this design, we relate the number of injected deviations in the high-anomaly stratum to direct-hit probability and then measure the cost of executing the resulting audit budgets. In real 200-step 7B traces, the high stratum captures 9 of 15 learning-rate deviations across three seeds, but only 1 of 15 skipped updates. This difference matters in practice: on a sealed 1.5B trajectory, budgets calibrated offline to a 90\% direct-hit target reduce median audit time by 49\% and 67\% for two learning-rate deviation settings. With no high-stratum capture, the five-deviation setting instead costs 38\% more than uniform sampling. These comparisons use known attack labels for budget calibration and measure complete audits of the same honest trajectory, separating selection effects from early rejection.}

\revzk{The system verifies selected intervals using disclosed states and batches with deterministic BF16 replay. A 7B integrated run demonstrates sealing, random challenges, independent replay, and rejection; separate experiments reproduce 2{,}100 steps bit-exactly across 7B--14B models. Together, these results connect the sampling analysis to an implemented audit and establish the numerical basis for checking the selected computation.}

\revzk{In summary, we make three contributions:}
\begin{itemize}[leftmargin=*]
\item \revzk{\textbf{An integrated audit path for sealed training records.} We integrate post-seal challenges, complete-state restoration, and independent interval replay to check selected state transitions and their reported telemetry (Section~\ref{sec:protocol}).}
\item \revzk{\textbf{Telemetry-guided sampling with replay-cost analysis.} We combine anomaly-stratified and baseline random sampling, characterize their conditional hit probabilities, and account for the replay work needed to execute the checks (Section~\ref{sec:model}).}
\item \revzk{\textbf{Measured audit savings and their conditions.} Real fine-tuning experiments establish integrated 7B auditing and 49\%/67\% wall-time savings on 1.5B under offline-calibrated budgets. The skipped-update result identifies when preferential sampling increases cost; component measurements explain replay fidelity and recording overhead (Section~\ref{sec:evaluation}).}
\end{itemize}

\section{Related Work}
\label{sec:related}

\revzk{\method{} builds on selective training verification, deterministic replay, and cryptographic commitments. Table~\ref{tab:sota-comparison} summarizes the relevant verification mechanisms. Our focus is the interaction between telemetry-guided selection and the cost of independently checking the selected intervals.}

\begin{table*}[t]
\centering
\caption{\revzk{Training-verification mechanisms, selection strategies, and trust assumptions.}}
\label{tab:sota-comparison}
\small
{\color{black}
\begin{tabularx}{\textwidth}{@{}p{3.8cm}XX@{}}
\toprule
Approach & Selection and verification & Relevant scope or trust assumption \\
\midrule
Per-step ZK~\cite{verilora,vft} & Proofs of encoded training relations & Witness privacy; soundness depends on the complete encoded relation \\
VeriML~\cite{veriml} & Commit iterations, prove random selections & Sampled cryptographic verification \\
Proof-of-Learning~\cite{pol} & Prioritize large updates, replay intervals & Recorded training states; numerical comparison \\
AFTUNE~\cite{aftune} & TEE-assisted sampled checking and sparse-checkpoint replay & Enclave trust; targeted selection is also considered \\
\method{} & Telemetry-stratified targets, independent interval replay & Opened states and batches; conditional sampling and replay costs \\
\bottomrule
\end{tabularx}}
\end{table*}

\subsection{Selective Training Verification}
\label{sec:rw-commitsample}
\revzk{VeriML~\cite{veriml} combines iteration commitments with proofs of randomly selected training computations. zkMLaaS~\cite{zkmlaas} uses challenge--response protocols over weight updates, while PoL-BFL~\cite{polbfl} combines sampled verification with a federated ledger. These systems precede our use of commit-then-sample verification. Proof-of-Learning~\cite{pol} records intermediate states and prioritizes large parameter updates for recomputation, establishing a direct precedent for signal-guided replay. Its subsequent security analysis~\cite{polbroken} also motivates care about what replay evidence actually establishes. \method{} develops this direction through sealed step telemetry, stratified random sampling, and measured replay costs at conditional hit budgets.}

\subsection{Hardware-Assisted Auditing and Replay}
\label{sec:rw-tee}
\revzk{AFTUNE~\cite{aftune} combines TEE-assisted verification with sampled recomputation, including sparse checkpoints, reconstruction from preceding checkpoints, and targeted sampling. Building on selective replay, we study how sealed loss and update-norm rankings affect the budgets and executed replay costs of disclosed-state fine-tuning audits. Verde~\cite{verde}, OVT~\cite{ovt}, and OPEN-1B~\cite{open1b} develop deterministic execution and optimistic verification under their respective arbitration assumptions. Hawkeye~\cite{hawkeye2026} reproduces NVIDIA GPU matrix multiplications exactly on a CPU, addressing a numerical building block of heterogeneous verification. We use deterministic replay as a numerical substrate for checking selected training intervals.}

\subsection{Private Training Proofs}
\label{sec:rw-perstep}
\label{sec:rw-crypto}
\revzk{VeriLoRA~\cite{verilora} and VFT~\cite{vft} construct proofs for LoRA fine-tuning, addressing private verification of encoded training computation. zkLLMPoT~\cite{zkllmpot} studies final-checkpoint properties. \method{} instead studies disclosed-state audits with probabilistic coverage of selected intervals. Extending this audit to private verification requires a complete interval relation and sealed bindings; Appendix~\ref{app:circuits} discusses the corresponding proof obligations and possible backend techniques.}

\subsection{Selective Inference Verification}
\label{sec:rw-inference}
\revzk{SLP~\cite{slp2026} commits chunk-boundary activations before challenges and proves selected chunks with input/output anchors. NanoZK~\cite{nanozk2026} links layerwise proofs and uses Fisher information for audit-budget triage, while full soundness requires every layer. Such work shares the use of selective computation checking. Training additionally carries optimizer, scheduler, and RNG state across steps; the audit must reconstruct these dependencies before it can check a selected transition.}

\section{\revzk{Audit Model and Sampling Analysis}}
\label{sec:model}

\subsection{Outsourced Fine-Tuning Setting and Participant Roles}
\label{sec:model-setting}

\revzk{A \emph{client} supplies dataset $D$ and training specification $\sigma$; an untrusted \emph{provider} performs fine-tuning; and an \emph{auditor}, either the client or its delegate, checks selected computation against sealed records. The provider has the plaintext data and states. The auditor is permitted to inspect the model, checkpoints, and batches needed for independent replay, together with commitments, telemetry, and challenge evidence. We study this disclosed-state audit relationship; requirements for a future private-proof backend are discussed in Appendix~\ref{app:circuits}.}

We use LoRA fine-tuning~\cite{hu2022lora}, with frozen base weights $\mathbf{W}_0$ and adapted weights $\mathbf{W} = \mathbf{W}_0 + (\alpha_{\mathrm{LoRA}}/r)\,\mathbf{B}\mathbf{A}$,
where $\mathbf{A}\in\mathbb{R}^{r\times d}$ and $\mathbf{B}\in\mathbb{R}^{d'\times r}$
with rank $r\ll d$. Training advances across $N$ discrete steps. Let
$S_i$ denote the complete training state at step $i$: LoRA weights
$\mathbf{W}_{\mathrm{LoRA}}^{(i)}$, AdamW first and second moments
$(m_i,v_i)$~\cite{adamw}, pseudo-random number generator (RNG) state $\rho_i$,
scheduler state, optimizer step counters, and data cursor. Each checkpoint
stores this complete state. Each step executes a transition
\begingroup
\setlength{\abovedisplayskip}{5pt}
\setlength{\belowdisplayskip}{5pt}
\setlength{\abovedisplayshortskip}{5pt}
\setlength{\belowdisplayshortskip}{5pt}
\begin{equation}
  T_i:\;(S_{i-1},\,x_i)\rightarrow S_i ,
  \label{eq:transition}
\end{equation}
\endgroup
\revzk{where $x_i$ is the mini-batch from committed dataset $D$. Under the
committed specification, Eq.~\eqref{eq:transition} makes the replay boundary
explicit: $S_{i-1}$ and $x_i$ are the inputs, and $S_i$ is the expected output.
A model-only snapshot is insufficient. Omitting optimizer moments, scheduler
position, RNG state, or the data cursor can change the executed
trajectory even when its visible weights appear plausible. We therefore treat
each checkpoint as a recoverable execution state rather than merely a saved
model. Auditing a selected step restores the nearest preceding sealed
checkpoint and replays the ordered transitions through that step, checking both
the target record and the state from which it was produced.}

\subsection{Adversarial Attack Space}
\label{sec:model-attacks}

\paragraph{Threat Model.} The provider is computationally bounded and malicious,
controlling execution end-to-end. It may attempt seven attack families:
\textbf{A1 Step skipping}; \textbf{A2 Compute reduction} (shrinking batches or
substituting a cheaper base model); \textbf{A3 Artifact forgery};
\textbf{A4 Transition tampering} (injecting targeted gradient perturbations or backdoors);
\textbf{A5 Randomness manipulation}; \textbf{A6 Specification falsification}
(altering learning rates or injecting uncommitted data);
and \textbf{A7 Seal-integrity violation} (rewriting training history post-hoc).
\revzk{The provider cannot break the hash binding or predict the post-seal challenge randomness. The reference public-beacon protocol additionally relies on the beacon's threshold trust assumptions. These attack families define the threat space; they are not all demonstrated in live LLM runs.}
\emph{Out of scope:} committed-but-harmful data poisoning (measured as a negative control in
Section~\ref{sec:eval-attack}), provider-auditor collusion, and side-channel leakage.

\paragraph{Verifier-Relative Violation Sets.} A corrupted state can propagate
numerically without invalidating later local transitions, because each later step
may be computed correctly from the altered predecessor. We therefore define the
violation set $V$ relative to the verification predicate: interval replay rejects
any sample whose replay path crosses an invalid transition, whereas a local-step
verifier rejects only the invalid transition itself and must not be credited with
the larger set. Consequently, the number of injected steps $k_{\mathrm{inj}}$ need
not equal $|V|$.

\subsection{Boundary Record Commitment Structure}
\label{sec:model-commitment}

During step $i\in\{1,\dots,N\}$, the provider samples a salt
$\omega_i \xleftarrow{\$} \{0,1\}^{256}$ and emits the commitment
\begin{equation}
  c_i = \commit(B_i;\,\omega_i) = \mathrm{BLAKE3}(B_i\,\|\,\omega_i).
  \label{eq:commit}
\end{equation}
\revzk{Table~\ref{tab:bi} in Appendix~\ref{app:bi_schema} gives the reference binding schema and its telemetry extension. Hash links fix record order; state, specification, batch, and final-weight digests bind the corresponding openings. Their consistency must still be checked by replay: a valid hash alone does not certify execution or the history before a checkpoint. The telemetry extension binds reported loss, parameter-update norm, and learning rate. The BF16 prototype implements this extension and binds full optimizer/RNG state. It checks batch hashes and positions; Section~\ref{sec:protocol} distinguishes this implementation from the reference membership relation.}

\subsection{Detection Probability Model}
\label{sec:model-detection}

\begin{proposition}[Uniform sampling]
\label{prop:zkspot-security}
Let $N$ be the number of sealed steps and $V\subseteq\{1,\dots,N\}$ with $|V|=k$ the fixed set of units rejected by the declared audit predicate. With $V$ fixed before independent post-seal randomness is drawn, uniform sampling of $s$ steps without replacement yields the exact rejection probability
\begin{equation}
  \Pdetect = 1 - \frac{\binom{N-k}{s}}{\binom{N}{s}} = 1 - \prod_{j=0}^{s-1} \frac{N-k-j}{N-j},
  \label{eq:hypgeom}
\end{equation}
or, under Bernoulli sampling with step inclusion rate $p_{\mathrm{inc}}$,
\begin{equation}
  \Pdetect = 1 - (1-p_{\mathrm{inc}})^k.
  \label{eq:bernoulli}
\end{equation}
\end{proposition}
\revzk{At $N{=}200$ and $s{=}10$, Eq.~\eqref{eq:hypgeom} gives rejection probabilities of 5.00\%, 22.83\%, and 40.85\% for $|V|{=}1,5,10$. To analyze \emph{direct hits}, we instead take the marked set to be the injected steps and substitute $k_{\mathrm{inj}}$ for $k$; we denote that event's probability by $P_{\mathrm{hit}}$. Direct hits, interval coverage, and actual rejection can differ because replay may check preceding steps. Appendix~\ref{app:injection_grid} gives statistical calibration and the separate conditional-deterrence model.}

\subsection{Telemetry Scoring and Stratified Sampling}
\label{sec:model-stratified}
\revzk{We rank sealed records using reported losses and parameter-update norms. Let $\mathcal{L}_i$ be the reported step loss and $u_i$ the reported norm of the parameter update. The loss-change feature is $d_i=|\mathcal{L}_i-\mathcal{L}_{i-1}|$ for $i>1$, with $d_1=0$. For each feature vector $v$, we compute its median $m(v)$ and scale $b(v)=1.4826\,\mathrm{median}_j|v_j-m(v)|$, then score}
\begin{equation}
 z_i(v)=\frac{|v_i-m(v)|}{b(v)},\qquad a_i=z_i(d)+z_i(u).
 \label{eq:telemetry-score}
\end{equation}
\revzk{If the scale is zero, we use the feature range; constant features contribute zero. The real-telemetry analysis uses these two equally weighted features. The BF16 audit also includes data-position increments, using the same normalization; Appendix~\ref{app:algorithm} specifies their initialization and missing-feature handling. We sort by decreasing $a_i$, breaking ties by increasing step index. This deterministic ranking selects where to check; the scores themselves do not determine acceptance.}

With $0<q\le1$ and $0\le\alpha\le1$, the auditor splits the budget $s\le N$ into two draws: the top-$q$ fraction of steps forms
a high-anomaly stratum $\mathcal{H}$ ($|\mathcal{H}|=\lceil qN\rceil$) receiving
$s_H=\min(|\mathcal{H}|,\lceil(1-\alpha)s\rceil)$ samples, while the remaining
$s_R=s-s_H$ samples are drawn uniformly from all records not selected in the first draw, including unselected members of $\mathcal{H}$. For a fixed marked set of size $k$, let $k_H$ count its members in $\mathcal{H}$. The probability of selecting at least one marked step is
\begin{equation}
  P_{\mathrm{hit}} = 1-\frac{\binom{|\mathcal{H}|-k_H}{s_H}}{\binom{|\mathcal{H}|}{s_H}}
  \cdot\frac{\binom{N-s_H-k}{s_R}}{\binom{N-s_H}{s_R}},
  \label{eq:stratified}
\end{equation}
\revzk{In the telemetry experiments, the marked set comprises the injected steps, so $k=k_{\mathrm{inj}}$. Taking the marked set to be $V$ instead gives $\Pdetect$. The second factor conditions on the first draw missing every marked step; if that draw must hit, the probability is one without evaluating the second factor. At $\alpha{=}1$, the rule recovers uniform sampling. Sealing fixes the ranking before the draw but cannot prevent deliberately smoothed reports. When $k_H{=}0$, only the baseline draw can hit; rounding can even make $s_R{=}0$ at small budgets. Thus preferential sampling benefits from measured capture, not from a presumed positive coverage floor.}

\subsection{\revzk{From Sample Budgets to Replay Cost}}
\label{sec:model-cost}
\revzk{For fixed $k$ and $k_H$, we enumerate $s=1,\dots,N$ to find the smallest budget meeting the direct-hit target in Eq.~\eqref{eq:stratified}. These counts require labeled deviations and are unknown in deployment. Each target $i$ requires replay from checkpoint $\kappa(i)=g\lfloor(i-1)/g\rfloor$, possibly with additional predecessor intervals for telemetry checks. A serial cost model uses startup cost $t_{\mathrm{start}}$, total restore time $t_{\mathrm{restore},i}$, actual replay count $r_i$, step cost $t_{\mathrm{step}}$, and residual checking cost $t_{\mathrm{check},i}$:}
\begin{equation}
 \revzk{C(Q)=t_{\mathrm{start}}+\sum_{i\in Q}
 \bigl(t_{\mathrm{restore},i}+r_i t_{\mathrm{step}}+t_{\mathrm{check},i}\bigr).}
 \label{eq:audit-cost}
\end{equation}
\revzk{Here $r_i$ includes repeated work and predecessor checks, and $t_{\mathrm{restore},i}$ includes their additional restores. Deduplicating intervals can undercount a verifier's work when it checks targets separately. We therefore report complete wall times alongside cost projections, distinguishing complete honest audits from runs that stop at the first rejection.}

\section{\revzk{\method{} Design}}
\label{sec:protocol}

\method{} coordinates outsourced LLM fine-tuning across five ordered phases:
\emph{pre-commit}, \emph{train-and-commit}, \emph{seal}, \emph{challenge}, and
\emph{recompute-and-verify}, as illustrated in Figure~\ref{fig:framework}. The design
\revzk{separates continuous recording from selective verification, with asynchronous I/O within the recording phase. We describe the reference protocol using a public beacon and dataset Merkle root. The evaluated BF16 implementation follows the same seal--sample--replay flow using local post-seal CSPRNG randomness and batch hashes/positions; it does not implement public-beacon timing or dataset membership proofs. An auxiliary integer checker evaluates a separate computation relation.}

\subsection{Commit--Seal--Sample--Verify Protocol}
\label{sec:protocol-design}

\paragraph{Phase 1: Pre-Commitment.} Prior to execution, the client and provider
bind the global specification
\begin{equation}
  \sigma = (\mathrm{base\_id}, \mathrm{rev}, C_{\mathrm{base}}, \mathrm{LoRA\_cfg}, \mathrm{hp}),
  \label{eq:sigma}
\end{equation}
where $C_{\mathrm{base}}$ commits to the frozen base weights and $\mathrm{hp}$ denotes
the hyperparameters; the fingerprint $\sigid = h(\sigma)$ is embedded into every step's
boundary record. The client commits dataset $D$ via a Merkle root $R_D$, and the
public metadata $(\sigma, \sigid, R_D)$ is signed and published. The provider stores the complete initial checkpoint $S_0$ and binds its digest to the first record.

\paragraph{Phase 2: Train-and-Commit (Always-On).} During step $i\in\{1,\dots,N\}$, the
provider computes $T_i: (S_{i-1}, x_i) \to S_i$ under deterministic floating-point
rules (Section~\ref{sec:protocol-numerics}) and emits a salted BLAKE3 commitment per
Eq.~\eqref{eq:commit}. Every $g$ steps, the full state $\hat{S}_{jg}$ is
cached as a sealed checkpoint. A pinned-memory double buffer and dedicated writer thread overlap serialization with execution, reducing measured checkpoint stalls (Section~\ref{sec:eval-commit}). Setting $g{=}1$ reduces the interval
predicate $\kappa = g\lfloor(i-1)/g\rfloor$ to single-step predecessor checking $\kappa = i-1$,
which we treat as a special case of the protocol rather than an external baseline.

\paragraph{Phase 3: Seal.} Upon completing step $N$, the provider exports an atomic
seal record
\begin{equation}
  \begin{split}
  \mathrm{Seal} = \{&\sigid, R_D, C_{\mathrm{base}}, N, g, R_{\mathrm{chain}}, \\
  &R_{\mathrm{merkle}}, \mathrm{strat}, \mathrm{bcn}, h(\mathbf{W}_{\mathrm{final}}), T_{\mathrm{seal}}\},
  \end{split}
  \label{eq:seal}
\end{equation}
Here $R_{\mathrm{chain}}=c_N$ and $R_{\mathrm{merkle}}=\mathrm{Merkle}(c_1,\dots,c_N)$.
The descriptor $\mathrm{strat}$ binds the scoring rule, feature weights, $q$, $\alpha$,
and budget $s$; $\mathrm{bcn}$ identifies the beacon and round-selection rule.
These parameters are registered before training.
The auditor validates the chain and acknowledges the seal ($\textsf{seal\_ack}$),
freezing timestamp $T_{\mathrm{seal}}$. Once sealed, any modification of steps
alters $R_{\mathrm{chain}}$ or $R_{\mathrm{merkle}}$, which fixes the violation set $V$
before the challenge is revealed.

\paragraph{Phase 4: Post-Seal Public Challenge.} Following $\textsf{seal\_ack}$,
the auditor fetches the beacon value $b_t$ of the earliest drand quicknet round
emitted at $t > T_{\mathrm{seal}}$ and derives the challenge seed
\begin{equation}
  \xi = h\bigl(b_t\,\|\,R_{\mathrm{chain}}\bigr).
  \label{eq:challenge}
\end{equation}
It then draws the challenge set $Q \subseteq \{1,\dots,N\}$ ($|Q|{=}s$) under the
seeded two-stratum sampling rule of Eq.~\eqref{eq:stratified}: the top-$q$ anomaly
stratum $\mathcal{H}$, ranked by sealed step telemetry, contributes
$s_H=\min(|\mathcal{H}|,\lceil(1-\alpha)s\rceil)$ samples via
$\mathrm{PRF}(\xi\,\|\,\text{``high''})$, and the remaining $s_R=s-s_H$ samples are
drawn uniformly \revzk{from all records not already selected} via $\mathrm{PRF}(\xi\,\|\,\text{``base''})$;
$\alpha{=}1$ recovers uniform sampling exactly. Coupling threshold beacon randomness
to $R_{\mathrm{chain}}$ \revzk{binds the challenge to the acknowledged seal, provided the provider cannot choose among multiple seals or beacon rounds after observing their randomness.}

\paragraph{Phase 5: Recompute-and-Verify.} Each challenged step $i\in Q$, together
with its replay interval from checkpoint $\kappa=g\lfloor(i-1)/g\rfloor$, forms an
\emph{audit unit}. The provider opens the checkpoint, required batches, and sealed interval records. The auditor checks the checkpoint digest, restores its complete state, and independently re-executes the $i-\kappa$ steps, comparing state and telemetry bindings at each transition. To validate a target's loss change, it also verifies the preceding loss; when $i-1=\kappa$, this requires replaying the previous interval. Missing openings and mismatches produce explicit rejection evidence. The BF16 cost measurements include the auditor's loading, restoration, replay, and checks. Provider-side proving and parallel scheduling belong to the separate component workload in Appendix~\ref{app:ledger}. Algorithm~\ref{alg:zkspot} in Appendix~\ref{app:algorithm} specifies the reference flow and its failure conditions.

\subsection{\revzk{Independent Replay and Auxiliary Checking}}
\label{sec:protocol-circuits}

\revzk{The \textbf{BF16 replay backend} checks audit units on a matching accelerator type and software stack. BoundaryUnit v2 binds reported loss and parameter-update norms, batches, and complete checkpoints; the verifier recomputes these values with each transition. Separate 7B--14B experiments establish bit-exact interval replay in the tested environments (Section~\ref{sec:eval-numerics}).}

\revzk{The \textbf{auxiliary integer checker}, \texttt{nfzk}, opens state and batch values, independently replays the encoded step, and checks exact GEMM, quantization, AdamW, and binding constraints. Its tiny-$\sigma$ fixture has two layers, width 64, sequence length 32, batch 2, LoRA rank 8, and 424{,}014 scalar constraints. Separate Freivalds microbenchmarks check integer product claims across CPU/GPU hardware (Appendix~\ref{sec:eval-freivalds}). The component analysis uses a replay-only model and an integer-checker model, labeled Replay and Integer in tables. These price separate workloads. Appendix~\ref{app:circuits} specifies the obligations of a future private backend.}

\subsection{Deterministic Bit-Exact Numerics}
\label{sec:protocol-numerics}

Heuristic tolerance comparators define a metric ball of acceptable outputs, allowing
adversaries to accumulate stealthy model shifts~\cite{polbroken}. \method{} instead
enforces \emph{bit-exact} execution between training and recomputation. Integer
\revzk{fixed-point arithmetic supports reproducibility across CPU and GPU hardware, but our retained probes expose overflow and dynamic-range problems in transformer execution (Section~\ref{sec:eval-numerics}).} \method{} therefore
\revzk{uses deterministic floating-point execution, following the replay principle explored by prior systems~\cite{verde,open1b}. Deterministic kernel settings, restored RNG state, and complete optimizer and scheduler checkpoints enable bit-exact recomputation in the tested matching environments. The measured scope and overhead appear in Sections~\ref{sec:eval-commit} and~\ref{sec:eval-numerics}.}

\section{Empirical Evaluation}
\label{sec:evaluation}

\revzk{We first evaluate whether sealed training records support independent audit decisions, then examine when telemetry improves sampling. We then compare measured audit costs at offline-calibrated direct-hit targets. Component measurements establish replay fidelity and online overhead. We distinguish cost projections at additional operating points from complete system runs.}

\subsection{Setup and Provenance}
\label{sec:eval-setup}

\textbf{Workload.} The primary workload fine-tunes Qwen2.5-1.5B~\cite{qwen25} via LoRA~\cite{hu2022lora}
($r{=}8, \alpha_{\mathrm{LoRA}}{=}16$) on Databricks dolly-15k (revision \texttt{bdd27f4})~\cite{dolly}:
micro-batch 8, 4-step gradient accumulation (effective batch 32),
sequence length 512, AdamW ($2\times 10^{-4}$, cosine decay)~\cite{adamw},
deterministic 90/10 split, seeds $\{42, 1024, 2048\}$.
Scale and model-family coverage comes from 300-step $\times$ 3-seed runs on Qwen2.5-7B
and Llama-3.1-8B-Instruct~\cite{llama31}, plus a Qwen2.5-14B capacity probe. We
compare native BF16 kernels (\emph{Native}) against our deterministic floating-point
configuration (\emph{DetFloat}; Section~\ref{sec:protocol-numerics}).

\textbf{Hardware Infrastructure.} End-to-end training runs execute on NVIDIA
RTX-6000D GPUs (PyTorch 2.13.0+cu130, Transformers 5.12.1); prover micro-benchmarks
and the multi-GPU component workload run on NVIDIA A100-40GB PCIe hosts. \revzk{The 1.5B cost comparison uses five A100 GPUs for concurrent one-GPU jobs; shared CPU and I/O resources can affect wall time.}

\revzk{\textbf{Measurement scope.} Table~\ref{tab:evidence} distinguishes integrated audits, telemetry analysis, and supporting component measurements. We report measured wall time for complete BF16 audits and label calculations based on component timings as cost models. Appendices~\ref{app:cost-models} and~\ref{app:ledger} specify the inputs to those models.}

\begin{table}[t]
\centering
\caption{\revzk{Evidence paths and reported quantities.}}
\label{tab:evidence}
\footnotesize
{\color{black}
\begin{tabularx}{\columnwidth}{@{}p{.27\columnwidth}X@{}}
\toprule
Evidence & Measured object or calculation \\
\midrule
7B BF16 audit & One 100-step sealed trajectory; independent sampled replay and rejection \\
1.5B audit costs & 29 complete honest audits and nine attack-rejection runs; offline budget calibration \\
Real telemetry & 48 runs at $N{=}100$ and six at $N{=}200$; ranking measured, hit probabilities calculated \\
7B--14B replay & 2{,}100 bit-exact replayed steps \\
Integer checker & Tiny-$\sigma$ relation; separate $d{=}4096$ product microbenchmark \\
Sampling/ledger & Simulated telemetry and component-priced costs; separate two-layer GPU scaling \\
\bottomrule
\end{tabularx}}
\end{table}

\begin{figure*}[t]
\centering
\includegraphics[width=\textwidth]{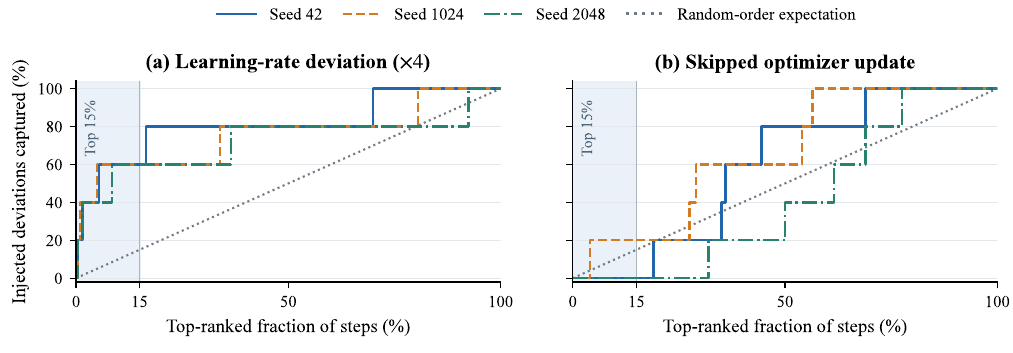}
\caption{\textbf{Telemetry rankings concentrate learning-rate deviations more consistently than skipped updates.} Each curve uses one real 7B trajectory ($N{=}200$, five injected steps). We rank all steps by sealed loss/update-norm scores and report the fraction of injected steps within each ranked prefix. The shaded region is the top 15\% used by the sampler; the dotted diagonal is the expectation under random ordering.}
\Description{Two panels show cumulative injected-step capture for learning-rate deviations and skipped updates across three seeds. At the top 15 percent, learning-rate capture is 60 percent for every seed, while skip-update capture is zero, 20 percent, and zero.}
\label{fig:telemetry-capture}
\end{figure*}

\subsection{\revzk{Independent Auditing on a Sealed 7B Trajectory}}
\label{sec:eval-cost}
\label{sec:eval-e2e}

\revzk{We fine-tune Qwen2.5-7B with BF16 LoRA for 100 steps and checkpoint every five steps. Before training, we fix a false loss report at step 19, a false update-norm report at step 35, and a skipped optimizer update at step 51. The skipped update retains forward/backward computation; we do not measure a provider cost saving. BoundaryUnit v2 commits reported telemetry and complete state. After sealing, local cryptographic randomness seeds the original sampler with $q{=}0.15$, $\alpha{=}0.2$, and $s{=}10$; a separately launched verifier checks the resulting intervals.}

\revzk{The random challenge contains ten targets. The verifier accepts the first seven and rejects target 35, whose reported update norm is zero while recomputation gives approximately 0.085163. It stops at that rejection after visiting eight targets and replaying 34 steps, including repeated and necessary predecessor intervals. Separate fixed-target diagnostics reject the loss report at step 19 and the skipped update at step 51; the random run does not reach step 51. An honest control passes. Appendix~\ref{app:real-audit} records the challenge and individual outcomes.}

\revzk{Training takes 363.4\,s within the training phase; the verifier takes about 145.0\,s internally, with an externally sampled wall-time upper estimate of 147.7\,s including loading and checks. These times describe this run on an RTX 6000D, with another service resident on the GPU. We evaluate matched direct-hit budgets separately in Section~\ref{sec:eval-measured}.}

\begin{figure*}[t]
\centering
\includegraphics[width=\textwidth]{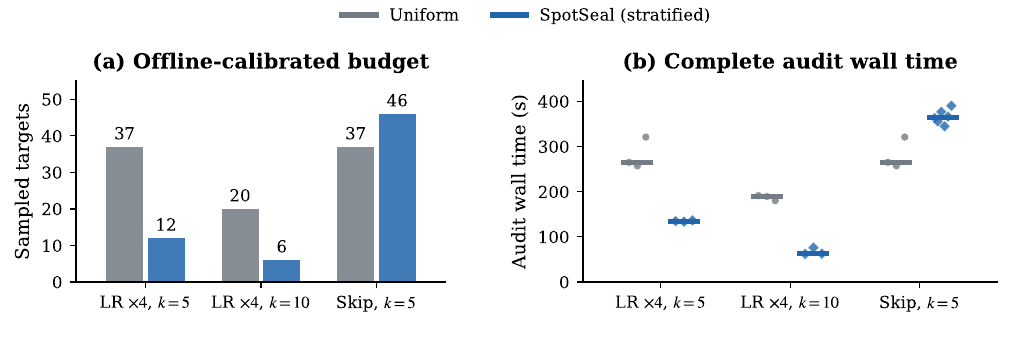}
\caption{\textbf{Offline-calibrated budgets and measured complete-audit costs.} We calibrate budgets to at least 90\% direct-hit probability using labeled learning-rate deviations (LR $\times4$) or skipped updates, then execute them on the same honest 1.5B trajectory. Dots show individual timings and horizontal marks show medians: three runs per arm, except six for stratified $k{=}5$ skipped updates. Both $k{=}5$ comparisons reuse the same uniform runs.}
\Description{Uniform and stratified budgets are 37 versus 12, 20 versus 6, and 37 versus 46. Complete-audit median time decreases by 49 and 67 percent in the learning-rate cases and increases by 38 percent in the skipped-update case.}
\label{fig:audit-cost}
\end{figure*}

\subsection{\revzk{Real Telemetry and Conditional Sampling Costs}}
\label{sec:eval-stratified}

\revzk{We measure loss and actual parameter-update norms in 48 real 100-step runs: 1.5B/7B models, skipped updates or learning rate multiplied by four, $k=k_{\mathrm{inj}}\in\{1,2,5,10\}$, and three seeds. Six 7B runs at $N{=}200$, $k{=}5$ repeat the two attacks across the same seeds. The sampler uses $q{=}15\%$ and $\alpha{=}0.2$ throughout. Figure~\ref{fig:telemetry-capture} presents the six 200-step rankings. The top 15\% captures 60\% of learning-rate deviations in every seed, versus 0\%, 20\%, and 0\% of skipped updates. The ranking therefore concentrates the former deviations more consistently.}

\revzk{For these 200-step trajectories, the conditional budget for 90\% direct-hit probability is 19 versus 74 uniform samples for learning-rate deviations in all three seeds. For skipped updates it is 92, 33, and 92, so stratification is worse in two runs. The injection positions are fixed across seeds, and these budgets use labels unavailable during deployment. Interval coverage is a different event: at $g{=}5$, $s{=}10$, skip-update coverage is 15.05\%, 54.89\%, and 37.71\%, versus 55.00\% for uniform. These calculations do not model the BF16 verifier's extra predecessor checks or stopping policy. Table~\ref{tab:real-telemetry} in Appendix~\ref{app:real-audit} gives exact capture and hit probabilities, alongside ranking quality and the telemetry-smoothing stress test.}

\begin{figure}[t]
\centering
\includegraphics[width=\columnwidth]{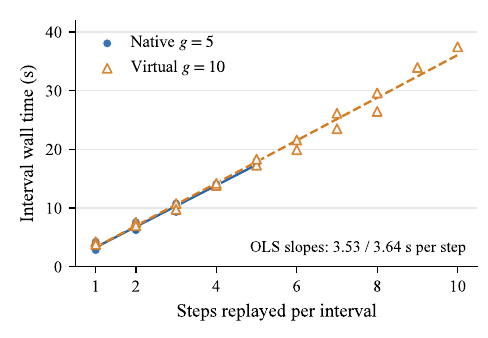}
\caption{\textbf{Interval replay time grows approximately linearly with replay depth on 7B.} Points are 13 successful native-$g{=}5$ and 18 successful virtual-$g{=}10$ intervals; lines are ordinary least-squares fits. The virtual stride uses existing checkpoints from the same $g{=}5$ trajectory. Two rejected intervals are excluded. Timings omit outer-verifier validation and extra predecessor replay; this is a primitive measurement, not a complete stride sweep.}
\Description{Interval wall time rises approximately linearly with replay depth, with fitted slopes of 3.53 seconds per step at native stride five and 3.64 at virtual stride ten.}
\label{fig:replay-cost}
\end{figure}

\subsection{\revzk{Measured Audit Cost at Matched Direct-Hit Targets}}
\label{sec:eval-measured}

\revzk{We next test whether reduced budgets save wall time in the implemented verifier. We train Qwen2.5-1.5B for $N{=}100$ steps with $g{=}5$ and training seed 42, retaining an honest trajectory and three deviation trajectories: learning rate multiplied by four at five or ten block-ending steps (V5L/V10L), and five skipped updates (V5S). Their high-stratum captures are respectively 2/5, 5/10, and 0/5. Using these labels offline, we choose the smallest budgets reaching at least 90\% direct-hit probability under Eq.~\eqref{eq:stratified}. We then measure complete audits at those budgets on the same sealed honest trajectory, avoiding differences caused by early rejection. Each run uses one A100-40GB GPU; concurrent jobs share host resources.}

\revzk{Figure~\ref{fig:audit-cost} presents the budgets calibrated from the attack rankings and their complete-audit costs on the honest trajectory. For learning-rate deviations, stratified budgets of 12 and 6 replace uniform budgets of 37 and 20, reducing median wall time by 49\% and 67\%. For five skipped updates, zero high-stratum capture increases the budget from 37 to 46 and median time by 38\%. Individual timings show run variability; both five-deviation comparisons share the same uniform baseline. These savings concern audit time under offline-calibrated budgets: $k$ and $k_H$ are unknown in deployment, and the honest trajectory supplies its own ranking.}

\revzk{Across the full cost matrix, all 29 honest-trajectory audits verify. Separately, stratified, uniform, and full-coverage audits each reject all three deviation trajectories, giving nine rejection runs with one challenge seed per arm. The actual verifier can also reject through additional predecessor checks, so the calibrated direct-hit target is not its complete interval-detection probability. Appendix~\ref{app:measured} gives the exact budgets and medians in Table~\ref{tab:measured-cost}, including the hypothetical zero-capture boundary, plus timing variability and rejection costs.}

\subsection{\revzk{What Determines Replay Cost?}}
\label{sec:eval-replay-cost}
\revzk{Figure~\ref{fig:replay-cost} shows interval timings on a retained 7B trajectory. Successful intervals give fitted replay costs of 3.53--3.64\,s per step, while six hot-cache checkpoint restores have a median of 0.106\,s. Replay therefore dominates the measured intervals. The longer intervals use virtual $g{=}10$ checkpoints selected from an existing $g{=}5$ trajectory; these primitive measurements omit the outer verifier's additional checks. Appendix~\ref{app:measured} records their scope.}

\revzk{Together, Figures~\ref{fig:audit-cost} and~\ref{fig:replay-cost} connect target allocation to executed work: reducing the number of targets helps only to the extent that their replay and checking costs also fall. Overlapping intervals and predecessor checks can repeat work, so sample counts alone cannot determine wall time. Appendix~\ref{app:cost-models} isolates these effects in controlled sampling and checkpoint models, and Figure~\ref{fig:frontier} presents their direct-hit--cost frontier. The separate component-priced proving ledger is documented in Appendix~\ref{app:ledger}.}

\subsection{Always-On Runtime Overhead: Four-Stage Breakdown}
\label{sec:eval-commit}

We decompose the DetFloat pipeline into four cumulative stages:
\textbf{S0} native BF16 kernels; \textbf{S1} deterministic kernels;
\textbf{S2} plus the per-step state digest $h(S_i)$; and \textbf{S3} plus the BLAKE3
commitment and checkpoint I/O, either synchronous or asynchronous. We measure 30
cells (2 models $\times$ 3 sequence lengths $\times$ 5 stage variants), each repeated
3 times, and Table~\ref{tab:e2b7} reports the results. Three findings emerge.
\revzk{\emph{First, recording costs are small}: state digests add at most 0.74\% and the commit/checkpoint stage at most 0.20\% marginal step time,} and Scheme A (salted BLAKE3,
0.079\,ms/step at 7B) remains the selected committer (Appendix~\ref{app:commit_scheme}).
\emph{Second, asynchronous double-buffering reduces checkpoint stalls}: per-checkpoint spikes
collapse from about 184\,ms to about 5.5\,ms (stall ratio $\le 1.2\times10^{-6}$
versus the 2\% bound). \emph{Third, determinism dominates}: deterministic kernels add
$+13.2\%$ to $+77.4\%$ (end-to-end $+13.4$--$25.9\%$ at 7B for sequence lengths
512--1024 and $+49.9\%$ at 2048), \revzk{a provider-side cost independent of audit budget $s$.}

\begin{table}[t]
\centering
\caption{\textbf{Marginal step-time cost of each pipeline stage} (median p50 delta vs.\ previous stage; S1 deterministic kernels, S2 state digest, S3 salted commit and async checkpoint; E2E = total vs.\ native BF16). Deterministic kernels dominate; each additional audit-specific stage contributes at most 0.74\%.}
\label{tab:e2b7}
\footnotesize
\setlength{\tabcolsep}{3.5pt}
\begin{tabular}{llcccc}
\toprule
\textbf{Model} & \textbf{Seq} & \textbf{S1 Det.} & \textbf{S2 Digest} & \textbf{S3 Commit} & \textbf{E2E} \\
\midrule
1.5B & 512  & $+17.3\%$ & $+0.74\%$ & $+0.20\%$ & $+18.4\%$ \\
1.5B & 1024 & $+36.6\%$ & $+0.59\%$ & $+0.13\%$ & $+37.6\%$ \\
1.5B & 2048 & $+77.4\%$ & $+0.64\%$ & $+0.02\%$ & $+78.6\%$ \\
7B   & 512  & $+13.2\%$ & $+0.35\%$ & $-0.15\%$ & $+13.4\%$ \\
7B   & 1024 & $+25.3\%$ & $+0.42\%$ & $+0.11\%$ & $+25.9\%$ \\
7B   & 2048 & $+49.5\%$ & $+0.26\%$ & $+0.02\%$ & $+49.9\%$ \\
\bottomrule
\end{tabular}
\end{table}

\subsection{Bit-Exactness Across Model Scale}
\label{sec:eval-numerics}

\revzk{Deterministic replay reproduces all 2{,}100 evaluated steps bit-exactly across the tested 7B--14B configurations. Table~\ref{tab:e6scale} summarizes replay fidelity and agreement with native-BF16 held-out next-token top-1 accuracy.} On Qwen2.5-7B and Llama-3.1-8B-Instruct, full 300-step
$\times$ 3-seed DetFloat runs match native-BF16 accuracy within
$|\Delta_{\mathrm{acc}}| \le 0.011$\,pp, and all 1{,}800
replayed steps reproduce sealed weights, checkpoints,
and commitment channels bit-exactly (max $|{\Delta}|{=}0.0$). A 14B capacity
probe (Qwen2.5-14B) sustains deterministic training at 5.48\,s/step within 64.1\,GiB
and replays all 300 steps across 60 checkpoint intervals bit-exactly. In these
runs, DetFloat step time carries a $+9$--$11\%$ determinism premium at sequence
length 512, while the 3-rep stage benchmark of Table~\ref{tab:e2b7} measures
$+13.2\%$ at 7B. Appendix~\ref{app:nf_ablation} gives the 1.5B validation results and numerical probes motivating deterministic floating-point execution.

\begin{table*}[t]
\centering
\caption{\textbf{Multi-model next-token accuracy and bitwise replay (300 steps $\times$ 3 seeds, dolly-15k).} DetFloat runs match native BF16 accuracy within 0.011\,pp and replay every step bit-exactly. $^\dagger$Qwen2.5-14B is a single-seed capacity probe (60 checkpoint intervals fully replayed).}
\label{tab:e6scale}
\footnotesize
{
\begin{tabular}{lccccc}
\toprule
\textbf{Model} & \textbf{Token acc.\ Native$\to$DetFloat} & $\mathbf{|\Delta_{\mathrm{acc}}|}$ & \textbf{Bitwise Replay} & \textbf{DetFloat Step} & \textbf{Peak Mem.} \\
\midrule
Qwen2.5-7B & $63.16{\to}63.16\%$ & 0.005\,pp & 900/900 (100\%) & 3.47\,s & 54.6\,GiB \\
Llama-3.1-8B & $62.53{\to}62.54\%$ & 0.011\,pp & 900/900 (100\%) & 3.80\,s & 58.6\,GiB \\
Qwen2.5-14B$^\dagger$ & probe & -- & 300/300 (100\%) & 5.48\,s & 64.1\,GiB \\
\bottomrule
\end{tabular}%
}
\end{table*}

\subsection{Supplementary Enforcement and Calibration Checks}
\label{sec:eval-attack}

\revzk{We additionally evaluate a simulated 1{,}260-configuration injection grid}
spanning six of the seven attack families (A1--A4, A6, A7) plus a committed-poison
negative control, across 10 sampling strategies, 3 random seeds,
and 500 trials per cell. Table~\ref{tab:e5} in Appendix~\ref{app:injection_grid}
reports the calibrated simulation ($N{=}200,s{=}10$), whose state-modifying
generator assigns cascading violation sets $|V|\in\{2,39,47\}$; these are
simulation predicates rather than a proof that every self-consistent state
corruption invalidates subsequent transitions. Reference probabilities use exact
formulas in 1{,}134 cells and Monte Carlo estimates in 126
(Appendix~\ref{app:injection_grid}). They fall within the empirical 95\% Wilson
intervals in \textbf{98.7\%} of cases (1{,}243/1{,}260); the 17 uncovered cases
have absolute gaps at most 0.03. An independent 20{,}000-trial calibration
agrees closely (0.3957 vs.\ 0.3950 theoretical). On clean traces, the
false-positive rate is exactly \textbf{0 / 15{,}000}, and all \revzk{simulated} audit-rejection
cases trigger.

\paragraph{Verifier-Scope Probe.}
A CPU fixture fixes
$Q=\{7,8,10,11,13\}$ on a 20-step lineage with a fabricated update at step
7 and checkpoint stride 5. On the canonical verifier, the interval
predicate rejects both forgery forms on the contaminated units
8 and 10: a witness omitting the interval opening fails
\textsf{PROOF\_INVALID} (the legacy code for invalid evidence; 2/2), and a forged witness carrying a real
interval fails \textsf{BINDING\_MISMATCH} under auditor-side replay (2/2);
all five honest controls verify. A prior verifier revision without the
predicate accepted the same two forged witnesses, which motivates defining
$V$ relative to the verifier (Section~\ref{sec:model-attacks}). Auditor-side
interval replay costs 0.25\,CPU-s/step on this fixture; the probe checks
enforcement and does not estimate detection rates.

\paragraph{\revzk{Integer-Checker Attack and Rejection Evidence.}}
In a 20-unit run ($g{=}5$), an
adversary sealed a fabricated step-7 update without backpropagation. A drand quicknet beacon
(round 2{,}398{,}884, BLS12-381-verified) drew $Q=\{6,7,13\}$: units 6 and 13 verified
bit-exactly (424{,}014 opened-witness constraints, about 2.1\,MB of evidence per
unit), step 7 was intercepted via \textsf{BINDING\_MISMATCH} on
\revzk{$\bind_{\mathrm{out}}$, and the run emitted a rejection certificate. This exercises evidence generation, not deployed escrow or settlement.} Binding ablations expose three failures: omitting $\sigid$ permits undetected base-model swaps,
omitting $\bind_{x_i}$ admits uncommitted batches, and omitting checkpoint
binding admits last-block checkpoint forgery.

\section{\revzk{Discussion and Limitations}}
\label{sec:discussion}

\revzk{\textbf{Selection quality and executed work jointly determine savings.} \method{} connects low-cost training records to independently checkable audit units. The learning-rate experiments show that concentrating deviations in the high stratum can reduce both the required budget and complete-audit wall time. The skipped-update results explain the opposite outcome: preferential allocation spends checks on an uninformative stratum. This leads to two practical evaluation criteria---how many deviations a ranking captures, and how much replay the resulting targets require. Reporting both exposes costs that a sample-count comparison would miss.}

\revzk{\textbf{Audit scope and deployment assumptions.} A successful audit establishes consistency of the checked intervals with sealed records, not correctness of the entire training history, provider compute expenditure, or final model quality. The auditor obtains the required states and batches, and the BF16 path requires matching hardware and software. Its local post-seal randomness does not supply a public beacon or independently attested seal time, and batch hashes do not replace dataset membership proofs. The reference protocol describes these additional mechanisms separately. Economic deterrence further requires enforceable penalties; no escrow deployment is evaluated.}

\revzk{\textbf{Evaluation scope and extensions.} The principal cost comparison uses one honest 1.5B, 100-step trajectory, with budgets calibrated from labeled attacks; repeated challenges do not supply independent training seeds or an online guarantee. The 7B evidence comprises an integrated audit and primitive measurements, not an equivalent cost matrix. Deterministic recording adds a measured 13.4--49.9\% total runtime premium at 7B across tested sequence lengths. Extending the results requires label-free budget calibration, longer trajectories, and evaluation under adaptive telemetry manipulation. Shared replay could reduce repeated interval work. A private backend would additionally need to prove the interval computation, bindings, and membership relations together; Appendix~\ref{app:circuits} states these obligations, and Table~\ref{tab:privacy} summarizes current disclosure.}

\section{Conclusion}
\label{sec:conclusion}

\revzk{\method{} connects telemetry-guided selection to independently verifiable training intervals for outsourced LLM fine-tuning. Sealed loss and update-norm signals allocate checks, while replay supplies the evidence for acceptance or rejection. The 7B prototype establishes the integrated audit path. On the same honest 1.5B trajectory, budgets calibrated offline to a common direct-hit target reduce median audit wall time by 49\% and 67\% for learning-rate deviations, but increase it by 38\% when skipped updates evade the high stratum. The results establish that selective-audit efficiency depends on both where checks are allocated and how much computation those checks require.}

\bibliographystyle{mlsys2025}
\bibliography{refs}

\section*{AI Usage Disclosure}

AI-assisted coding and writing tools supported research code development,
experiment execution, evidence checks, figure preparation, and manuscript
drafting and language polishing. The authors are responsible for the content,
results, and claims.

\clearpage
\appendix
\mbox{}\label{supplement:start}
\section{Boundary Record Field Schema}
\label{app:bi_schema}

\revzk{Table~\ref{tab:bi} lists the base binding fields and the telemetry extension used by the audit in Section~\ref{sec:model-commitment}. BoundaryUnit v2 serializes the reported loss, parameter-update norm, and learning rate in field $f10$; data position is bound in $f6$, and state digests cover the complete checkpoint. The BF16 implementation checks positions and actual batches, without the reference Merkle-membership relation. Its public development salts do not instantiate the private-salt assumption below.}

\begin{table}[!htbp]
\centering
\caption{\textbf{Field-level schema of boundary record $B_i$.} Base fields and the telemetry extension bind the specification, states, batch, record order, and supplied telemetry; the verifier must check the corresponding invariants.}
\label{tab:bi}
\footnotesize
\begin{tabularx}{\columnwidth}{llX}
\toprule
\textbf{Field} & \textbf{Identifier} & \textbf{Cryptographic Binding Content} \\
\midrule
$f1$ & \texttt{version\_id} & Protocol format version identifier \\
$f2$ & $\sigid$ & Re-quotation of spec fingerprint $h(\sigma)$ \\
$f3$ & $i$ & Step index in the execution sequence ($1\le i\le N$) \\
$f4$ & $\bind_{\mathrm{in}}$ & Hash commitment to input state $S_{i-1}$ \\
$f5$ & $\bind_{\mathrm{out}}$ & Hash commitment to output state $S_i$ \\
$f6$ & $\bind_{x_i}$ & Batch position and membership proof against $R_D$ \\
$f7$ & $\link_i$ & Hash link $h(c_{i-1})$; for $i{=}1$, binds $\sigid, R_D, h(S_0), C_{\mathrm{base}}$ \\
$f8$ & \texttt{flags} & Checkpoint indicator, boundary flags (\texttt{is\_ckpt}, \texttt{is\_last}) \\
$f9^\dagger$ & $\bind_{\mathrm{final}}$ & Final weight digest $h(\mathbf{W}_{\mathrm{final}})$; populated only at $i{=}N$ \\
$f10$ & \texttt{telemetry} & Presence flag and reported loss, parameter-update norm, and learning rate (v2 extension) \\
\bottomrule
\multicolumn{3}{@{}p{\columnwidth}@{}}{\footnotesize $\omega_i$: 256-bit private uniform salt; $^\dagger$optional field instantiated on final unit.}
\end{tabularx}
\end{table}

\section{End-to-End Protocol Algorithm}
\label{app:algorithm}

\revzk{Algorithm~\ref{alg:zkspot} specifies the reference protocol, including public randomness and dataset membership. The BF16 implementation replaces these with local post-seal CSPRNG challenges and batch hash/position checks. It checks targets in increasing order and stops at the first rejection. A verifier execution fault produces an inconclusive outcome rather than evidence of a provider violation.}

\subsection{Scoring Rule and Feature Availability}
The implementation in \path{zkspot/challenge/stratified.py} constructs loss-change, update-norm, and data-position-increment features. Loss change is zero at step 1; otherwise it requires both adjacent losses. The position increment uses the most recent available position, initialized to zero. Each feature is normalized as in Eq.~\eqref{eq:telemetry-score}, using only available values to compute its median and MAD; a zero MAD falls back to the range. Missing entries, absent features, and constant features contribute zero. Nonnegative feature weights default to one, giving
\begin{equation}
 a_i=w_d z_i(d)+w_u z_i(u)+w_p z_i(p),
 \label{eq:weighted-telemetry}
\end{equation}
where $p$ denotes the position-increment vector. Ranking uses the key $(-a_i,i)$, so ties favor the earlier step. The scoring rule, weights, $q$, and $\alpha$ are fixed before challenge randomness; the plan binds the scores and stratum membership.

The real-telemetry analysis supplies loss and update norm, leaving position absent. The BF16 audit supplies all three fields through its boundary-record view. Its data cursor starts at zero: for a fixed batch advance, the position-increment vector begins with zero and then contains the repeated advance. The MAD is zero, so range normalization adds one to the first step's score and zero thereafter. We preserve this initialization when describing the evaluated implementation; its score is not identical to the two-feature analysis at step 1. Replay compares the recorded loss, update norm, learning rate, batch, and complete state with recomputed values. At a checkpoint boundary, it also replays the predecessor's interval and checks that loss against its sealed record, validating both inputs to the target's loss difference.

\begin{algorithm}[!tp]
\caption{The \method{} Spot-Check Auditing Protocol}
\label{alg:zkspot}
\footnotesize
\begin{algorithmic}[1]
\REQUIRE Specification $\sigma$ (including $N$), dataset root $R_D$, budget $s$, stride $g$, scoring rule and weights, $q$, $\alpha$, beacon descriptor $\mathrm{bcn}$.
\ENSURE \textsf{VERIFIED}, a rejection code, or \textsf{INCONCLUSIVE} on verifier execution faults.
\STATE On a verifier execution fault, return \textsf{INCONCLUSIVE}; missing provider openings follow the rejection checks below.
\STATE \textbf{Phase 1 (Pre-Commit):} Register $\sigid \gets h(\sigma)$, $R_D$, and sampling parameters; store $S_0$ and bind its digest;
\STATE \textbf{Phase 2 (Train-and-Commit):}
\FOR{$i = 1$ \TO $N$}
  \STATE Compute bit-exact state $S_i \gets T_i(S_{i-1}, x_i)$;
  \STATE Form $B_i$ with the telemetry extension in Table~\ref{tab:bi}, binding supplied values and their presence flag; sample $\omega_i \xleftarrow{\$} \{0,1\}^{256}$;
  \STATE Emit $c_i \gets \mathrm{BLAKE3}(B_i \,\|\, \omega_i)$;
  \IF{$i \bmod g = 0$}
    \STATE Store sealed checkpoint $\hat{S}_i \gets S_i$ (async double-buffered write);
  \ENDIF
\ENDFOR
\STATE \textbf{Phase 3 (Seal):} Provider publishes $\mathrm{Seal}$; Auditor freezes $T_{\mathrm{seal}}$ upon $\textsf{seal\_ack}$;
\STATE \textbf{Phase 4 (Public Challenge):}
\STATE Wait for drand beacon value $b_t$ emitted at $t > T_{\mathrm{seal}}$;
\STATE Derive challenge seed $\xi \gets h(b_t \,\|\, R_{\mathrm{chain}})$;
\STATE \revzk{Compute $a_i$ via Eq.~\eqref{eq:weighted-telemetry}; sort by $(-a_i,i)$ and form the top-$q$ stratum $\mathcal{H}$;}
\STATE Draw $Q_H \subseteq \mathcal{H}$, $|Q_H| = \min(|\mathcal{H}|,\lceil(1{-}\alpha)s\rceil)$ via $\mathrm{PRF}(\xi\,\|\,\text{``high''})$;
\STATE Draw $Q_R$ uniformly from $\{1,\dots,N\}\setminus Q_H$, $|Q_R| = s - |Q_H|$ via $\mathrm{PRF}(\xi\,\|\,\text{``base''})$; set $Q = Q_H \cup Q_R$;
\STATE \textbf{Phase 5 (Recompute-and-Verify):}
\FORALL{$i \in Q$ in increasing order}
  \STATE Locate checkpoint $\kappa \gets g \cdot \lfloor (i-1)/g \rfloor$;
  \STATE Provider opens $\hat{S}_\kappa$, batches $x_{\kappa+1},\dots,x_i$ with Merkle paths, and all interval records $B_j$ with salts $\omega_j$;
  \IF{any interval opening is missing}
    \RETURN $\textsf{PROOF\_INVALID}$;
  \ENDIF
  \IF{$h(\hat{S}_\kappa)$ differs from the sealed checkpoint digest}
    \RETURN $\textsf{CHECKPOINT\_MISMATCH}$;
  \ENDIF
  \STATE \textbf{Auditor} independently replays $\hat{S}_\kappa \to \dots \to S_i$ under the declared numerical specification;
  \FOR{$j = \kappa{+}1$ \TO $i$}
    \IF{state, telemetry, batch membership, or record opening disagrees with sealed $B_j,c_j$}
      \RETURN $\textsf{BINDING\_MISMATCH}$;
    \ENDIF
  \ENDFOR
  \IF{$i>1$ and $i-1=\kappa$}
    \STATE Replay the interval ending at $i-1$ with the same opening, state, and telemetry checks;
    \IF{any predecessor check fails}
      \RETURN the corresponding rejection code;
    \ENDIF
  \ENDIF
  \STATE Both losses underlying the target's loss difference have now been checked (step 1 uses zero difference);
\ENDFOR
\RETURN $\textsf{VERIFIED}$;
\end{algorithmic}
\end{algorithm}

\section{\revzk{Real-Telemetry and BF16 Audit Evidence}}
\label{app:real-audit}

\revzk{The 100-step matrix has 48 trajectories: Qwen2.5-1.5B/7B, two deviations (skipped optimizer update or learning rate multiplied by four), $k\in\{1,2,5,10\}$, and seeds 42/1024/2048. Six 200-step 7B runs repeat both deviations at $k{=}5$ across those seeds. They measure actual losses and parameter-update norms; rank quality, direct-hit probabilities, and oracle budgets are calculated offline from known injection labels. The larger matrix is not a set of complete BF16 verifier runs.}

\begin{table}[t]
\centering
\caption{\revzk{Real 7B telemetry at $N{=}200$, $k{=}5$. Entries follow seeds 42/1024/2048. Hit probabilities are calculated from measured rankings and known injection labels for $s{=}10$; uniform gives 22.83\% in every run. They are not verifier detection rates.}}
\label{tab:real-telemetry}
\footnotesize
{\color{black}
\begin{tabularx}{\columnwidth}{@{}lXX@{}}
\toprule
Attack & High-stratum capture & Direct-hit probability \\
\midrule
Skip update & 0/5, 1/5, 0/5 & 5.15\%, 30.45\%, 5.15\% \\
Learning rate $\times4$ & 3/5, 3/5, 3/5 & 64.02\% in each run \\
\bottomrule
\end{tabularx}}
\end{table}

\revzk{At $N{=}200$, skip-update ranking AUC is 0.598, 0.665, and 0.421; learning-rate deviation AUC is 0.823, 0.767, and 0.729. A label-informed counterfactual replaces telemetry at the injected steps with smoothed values and re-seals the records: high-stratum capture becomes zero, and the conditional 90\% direct-hit budget is 92 versus uniform's 74. This stress test demonstrates how provider-authored telemetry can defeat ranking; it is not an online adaptive-attack experiment.}

\revzk{For skip-update interval coverage, we enumerate the $g{=}5$ interval predicate under the frozen rankings, obtaining 15.05\%, 54.89\%, and 37.71\% at $s{=}10$; uniform yields 55.00\%. The calculation assumes that covering an injected step causes rejection. It omits extra predecessor checks and early stopping in the BF16 verifier and therefore is neither a measured detection rate nor an audit-time estimate.}

\subsection{\revzk{Integrated 7B Audit}}

\revzk{The independent BF16 implementation uses Qwen2.5-7B revision \texttt{d149729}, Dolly-15k revision \texttt{bdd27f4}, LoRA rank 8 with q/k/v/o projections, sequence length 512, micro-batch 8, accumulation 4, AdamW, and a cosine schedule. Complete checkpoints bind trainable parameters, Adam moments and step counters, scheduler, CPU/CUDA RNG, and data cursor. BoundaryUnit v2 binds reported loss, parameter-update norm, learning rate, and batch hashes. A separately supplied seal root fixes the verifier's input trajectory; a local cryptographically secure pseudorandom generator (CSPRNG) supplies one draw after sealing.}

\revzk{The resulting challenge is
\begin{equation}
Q=\{1,5,6,7,10,11,34,35,51,74\}.
\label{eq:bf16-challenge}
\end{equation} The verifier stops at target 35 after seven successful targets, with 34 total replayed steps including repeated intervals and predecessor checks. Target 51 was selected by the uniform draw but was not visited. Independent fixed-target diagnostics establish its rejection, as Table~\ref{tab:bf16-verdicts} reports. Replaying the supplied random value reproduces the same $Q$; that reproduction is a diagnostic, not another fresh random trial.}

\begin{table}[t]
\centering
\caption{\revzk{BF16 audit outcomes on the 100-step 7B trajectory. Fixed-target runs diagnose individual deviations; only one challenge was drawn randomly.}}
\label{tab:bf16-verdicts}
\footnotesize
{\color{black}
\begin{tabularx}{\columnwidth}{@{}lX@{}}
\toprule
Audit & Observed outcome \\
\midrule
Honest target 1 & Passes \\
Random challenge & Stops at 35: update norm 0 reported versus 0.085163 recomputed \\
Fixed target 19 & Rejects loss 1.567494 versus 1.317494 recomputed \\
Fixed target 35 & Rejects false update norm \\
Fixed target 51 & Rejects skipped update: norm 0 versus 0.061800 recomputed \\
\bottomrule
\end{tabularx}}
\end{table}

\revzk{Training-phase wall time is 363.413\,s and the process-internal total is 382.115\,s, excluding interpreter startup. The final random verifier records about 145.0\,s internally; one-second external polling gives an upper wall-time estimate of 147.7\,s. Model loading, data checks, checkpoint reads, and replay are included. Training peak tensor allocation is about 54.6\,GiB; verifier process residency sampled with \texttt{nvidia-smi} peaks near 71.9\,GiB. These are different memory measures, and neither isolates committer overhead. Another service occupies about 11\,GiB. No GPU-compute-time or same-goal speedup is inferred from these wall times.}

\subsection{\revzk{Negative Cases and Component Timing}}

\revzk{At 1.5B, actual false loss/update reports, skipped updates, and a batch swap are rejected. Separately constructed copies of an honest trajectory modify predecessor loss, Adam first moments, or data position before recomputing the seal; predecessor replay, state comparison, and cursor checks reject them. They are constructed counterexamples, not additional training runs. Post-seal edits, missing/wrong checkpoints, and missing records also fail the relevant checks. These tests exercise enforcement paths; they do not prove the protocol.}

\revzk{Five tiny canonical-verifier units were timed with one warmup and three measurements each. Their median total CPU costs range from 1.96 to 3.16\,s, including binding, independent step/interval replay, \texttt{nfzk}, and input derivation. Component timers overlap within this total and must not be summed. JSON parsing, constructor, training, network, and external proving are excluded. This timing boundary differs from the 3.32\,CPU-s residual calibration and the 1.2\,CPU-s binding-check assumption used in the component models.}

\revzk{The archive identifies the source snapshot for each verdict. The main implementation reports 59 regression tests passed and one skipped; a separate worker implementation reports 288 passed and five skipped. These counts describe different snapshots and are not pooled. Source-version details and final-output retests are recorded with the artifact.}

\section{\revzk{Measured Audit-Cost and Replay-Primitive Details}}
\label{app:measured}

\revzk{The 1.5B comparison uses sequence length 512, micro-batch 8, four accumulation steps, LoRA rank 8, AdamW at $2\times10^{-4}$, a cosine schedule, $N{=}100$, $g{=}5$, and training seed 42. Learning-rate deviations multiply the rate by four at steps 20/40/60/80/100 (V5L) or every tenth step (V10L); V5S skips the five block-ending updates. We fix $q{=}0.15$ and $\alpha{=}0.2$. Attack labels calibrate budgets offline, while complete-audit costs are measured on the same honest store. Table~\ref{tab:measured-budgets} gives the resulting direct-hit probabilities. The implementation can replay preceding blocks to check loss differences, so these probabilities do not exhaust its rejection predicate.}

\begin{table}[t]
\centering
\caption{\revzk{Measured complete-audit wall time on the same honest 1.5B trajectory. Budgets reach at least 90\% direct-hit probability under offline calibration; $k_H/k$ comes from the indicated attack ranking. Medians use three runs, except six for the five-deviation zero-capture arm. $^\dagger$Hypothetical zero-capture boundary, without a corresponding ten-skip trajectory.}}
\label{tab:measured-cost}
\footnotesize
{\color{black}
\begin{tabular}{@{}lcccc@{}}
\toprule
Calibration & $s_{\rm uni}$ & $s_{\rm strat}$ & Wall time (s) & Change \\
 & & & uni / strat & \\
\midrule
V5L: $2/5$ & 37 & 12 & 265.1 / 134.8 & $-49\%$ \\
V10L: $5/10$ & 20 & 6 & 189.3 / 62.5 & $-67\%$ \\
V5S: $0/5$ & 37 & 46 & 265.1 / 364.8 & $+38\%$ \\
Boundary: $0/10^\dagger$ & 20 & 32 & 189.3 / 290.6 & $+54\%$ \\
\bottomrule
\end{tabular}}
\end{table}

\begin{table}[t]
\centering
\caption{\revzk{Offline budget calibration for the measured 1.5B audit. The zero-capture $k{=}10$ row is a hypothetical boundary.}}
\label{tab:measured-budgets}
\footnotesize
{\color{black}
\begin{tabular}{@{}lcccc@{}}
\toprule
Rule & $k$ & $k_H$ & $s$ & Direct-hit probability \\
\midrule
Uniform & 5 & -- & 37 & 0.906640 \\
Stratified (V5L) & 5 & 2 & 12 & 0.915106 \\
Stratified (V5S) & 5 & 0 & 46 & 0.903587 \\
Uniform & 10 & -- & 20 & 0.904884 \\
Stratified (V10L) & 10 & 5 & 6 & 0.924917 \\
Stratified boundary & 10 & 0 & 32 & 0.907083 \\
\bottomrule
\end{tabular}}
\end{table}

\revzk{Most groups contain three challenge-seed runs; the $s{=}46$ group contains six timings. These repetitions share one trained trajectory, not independent training seeds. Two full-coverage runs have a median of 766.116\,s and each replay 395 steps because the implementation checks targets separately, including predecessor intervals. This is neither an optimized single sequential replay nor an equal-guarantee comparator for sampled auditing.}

\revzk{Additional best-capture calibration arms assume all violations lie in the high stratum, requiring six samples for $k{=}5$ and two for $k{=}10$. They are sensitivity cases, not observed attack rankings. In the $s{=}2$ arm, wall times are 44.8, 245.2, and 250.3\,s (median 245.217\,s); concurrent CPU/I/O contention makes reporting only the fastest run misleading. GPU residency is not GPU compute time. These cases and full-coverage runs complete the 29-run cost matrix.}

\begin{table}[t]
\centering
\caption{\revzk{Time to first rejection on three deviation trajectories (seconds, rounded). Each cell is one challenge-seed run. All nine runs reject on recomputed update-norm inconsistency; the table does not estimate detection rates.}}
\label{tab:measured-rejections}
\footnotesize
{\color{black}
\begin{tabular}{@{}lccc@{}}
\toprule
Trajectory & Stratified & Uniform & Full coverage \\
\midrule
V5L & 87 & 109 & 167 \\
V10L & 50 & 83 & 113 \\
V5S & 161 & 118 & 171 \\
\bottomrule
\end{tabular}}
\end{table}

\revzk{Table~\ref{tab:measured-rejections} reports the separate rejection runs. They stop after 7/7/20 visited targets for V5L (stratified/uniform/full), 3/4/10 for V10L, and 20/9/20 for V5S. These stopping times differ from the complete honest audits in Table~\ref{tab:measured-cost}.}

\revzk{The 7B primitive scan uses a retained $g{=}5$ trajectory on an RTX 6000D. It records 13 successful native-$g{=}5$ intervals and 18 successful virtual-$g{=}10$ intervals, plus two rejected virtual intervals excluded from the successful-replay fit. Fitted slopes are 3.5258 and 3.6404\,s per replayed step. Six restores of a 60{,}648{,}342-byte checkpoint have a median of 0.106387\,s under a warm OS page cache. The scan omits outer environment validation and predecessor checks. It does not establish cold-storage performance or an end-to-end stride optimum; exploratory $g{=}1$ values are estimates.}

\section{Numeric Format Ablation}
\label{app:nf_ablation}

\revzk{Table~\ref{tab:e1} summarizes numerical probes and retained validation measurements. The 300-step 1.5B experiment evaluates 1{,}502 examples over three seeds using held-out next-token top-1 accuracy, with mean 0.6010994 and population standard deviation 0.0001435. Fixed-point probes document overflow and dynamic-range failures. Their matched pre-training BF16 reference log is unavailable, so they support numerical diagnostics rather than accuracy comparisons.}

\begin{table}[!htbp]
\centering
\caption{\revzk{Numerical evidence and its scope. Multi-model accuracy and bitwise replay are reported separately in Table~\ref{tab:e6scale}.}}
\label{tab:e1}
\footnotesize
{\color{black}
\begin{tabularx}{\columnwidth}{@{}lX@{}}
\toprule
Experiment & Retained evidence \\
\midrule
Fixed-point probes & Overflow and dynamic-range failures; matched BF16 accuracy differences omitted \\
Exact-rsqrt probe & Counterfactual numerical diagnostic, distinct from the ordinary fixed-point path \\
1.5B BF16, 300 steps & Top-1 token accuracy $0.60110\pm0.00014$ over three seeds \\
7B/8B DetFloat & Native-BF16 agreement within 0.011\,pp in the matched runs \\
\bottomrule
\end{tabularx}}
\end{table}

\section{Commitment Scheme Microbenchmark}
\label{app:commit_scheme}

We evaluate three candidate commitment schemes on a 7B-parameter LoRA workload
over 3 repetitions, and Table~\ref{tab:e2} reports the results. Scheme A (BLAKE3
with a uniform 256-bit salt) requires only 0.079\,ms per step, about $0.002\%$ of a
3.47\,s 7B step; its measured wall-clock delta of $+0.04\%$ is within run-to-run
noise, as Scheme C's $-0.63\%$ shows. Scheme A has the lowest per-step cost,
$49\times$ below Poseidon and $9.4\times$ below BaseFold, and its 184--216 byte
\revzk{inputs use the benchmarked v1 record format, excluding the later v2 telemetry extension. These measurements motivate its use as the committer; they do not establish private-circuit cost.}

\begin{table}[!htbp]
\centering
\caption{\textbf{Commitment scheme comparison on 7B LoRA workload.} Scheme A (BLAKE3 with salt) achieves microsecond latency with v1 record inputs; sizes exclude the v2 telemetry extension. Runtime overhead is the wall-clock delta (3 reps), noise-dominated below ${\sim}1\%$.}
\label{tab:e2}
\resizebox{\columnwidth}{!}{%
\begin{tabular}{lccc}
\toprule
\textbf{Commitment Scheme} & \textbf{Cost / Step} & \textbf{Runtime Overhead} & \textbf{Input Representation} \\
\midrule
\textbf{A: BLAKE3 + Salt} & \textbf{0.079\,ms} & $\mathbf{+0.04\%}$ & \revzk{\textbf{184--216\,B} v1 input} \\
B: Poseidon Field Hash & 3.846\,ms & $+4.13\%$ & 6 field elements \\
C: BaseFold + Merkle & 0.746\,ms & $-0.63\%$ & GF($2^8$) RS encoding (0.24\,MB/s) \\
\bottomrule
\end{tabular}%
}
\end{table}

\section{Checkpoint Storage versus Recomputation Pareto Front}
\label{app:checkpoint_pareto}

\revzk{Table~\ref{tab:app_pareto} reproduces the retained $N{=}200$ storage/recomputation model. Each checkpoint is 39{,}223{,}360 bytes; file counts include the initial state. Mean replay distance includes the target step. Recomputation share is recomputation divided by proving plus recomputation, not a fraction of training time. The evaluated strides are $g\in\{10,50,100,500\}$.}

\begin{table}[!htbp]
\centering
\caption{\revzk{Checkpoint storage and recomputation model ($N{=}200$). Storage is derived from state size; replay uses measured step-time inputs. The initial checkpoint is included.}}
\label{tab:app_pareto}
\footnotesize
{\color{black}
\begin{tabularx}{\columnwidth}{@{}ccXXX@{}}
\toprule
$g$ & Files & Storage (GiB) & Replay steps & Recomp. share \\
\midrule
10 & 21 & 0.76712 & 5.5 & 3.31\% \\
50 & 5 & 0.18265 & 25.5 & 13.69\% \\
100 & 3 & 0.10959 & 50.5 & 23.90\% \\
500 & 1 & 0.03653 & 100.5 & 38.46\% \\
\bottomrule
\end{tabularx}}
\end{table}

\section{\revzk{Controlled Sampling and Checkpoint Cost Models}}
\label{app:cost-models}
\revzk{To isolate the cost mechanism, we also simulate telemetry under controlled separability. Uniform sampling needs 180 of 200 units to directly hit one invalid unit with 90\% probability. With $g{=}50$ and per-sample replay charged without deduplication, the calibration ledger costs $1.19\times$ direct per-step proving; at $g{=}5$ with deduplicated replay, its modeled advantage approaches parity. We then simulate three anomaly responses and a traceless control, pricing deduplicated replay as
\begin{equation}
 C_{\mathrm{model}}=s\,t_{\mathrm{unit}}+
 \left|\bigcup_{i\in Q}(\kappa(i),i]\right|t_{\mathrm{step}}\gamma,
 \label{eq:dedup-cost}
\end{equation}
where $t_{\mathrm{unit}}$ is the non-replay cost per selected target and $\gamma{=}2$ accounts for provider and auditor replay. The replay-only model sets $t_{\mathrm{unit}}{=}0$, with step-time inputs of 3.46/3.81/5.80\,s for 7B/8B/14B. The integer-checker model uses $t_{\mathrm{step}}{=}0.2505$\,CPU-s and $t_{\mathrm{unit}}{=}3.32$\,CPU-s, a residual obtained by subtracting replay rather than independently timing a checker. Table~\ref{tab:stratified} presents the \texttt{batch\_poison} slice. Modeled gains require anomaly separation, which often fails for real skipped updates.}

\begin{table*}[t]
\centering
\caption{\revzk{\textbf{Conditional sampling-cost model with simulated telemetry} ($N{=}200$, $g{=}5$, \texttt{batch\_poison}). $s^*$ targets 90\% direct-hit probability; $\E|\bigcup|$ is deduplicated replay volume over 24 injection seeds. Ratios price a full $N$-step audit against sampled work; they are not wall-time measurements. The replay-only model uses $N/\E|\bigcup|$; the integer-checker model adds the tiny-$\sigma$ residual cost. Figure~\ref{fig:frontier} presents the target-probability sweep.}}
\label{tab:stratified}
\footnotesize
\begin{tabular}{cccccccc}
\toprule
$\mathbf{k}$ & $\mathbf{s^*}$ uniform & $\mathbf{s^*}$ strat. & $\mathbf{s^*}$ reduction & $\E|\bigcup|$ strat. & $\E|\bigcup|$ uniform & \revzk{\textbf{Replay cost ratio}} & \revzk{\textbf{Integer cost ratio}} \\
\midrule
1  & 180 & 33 & $5.5\times$ & 80  & 195 & $2.51\times$  & $5.11\times$ \\
2  & 137 & 24 & $5.7\times$ & 57  & 182 & $3.53\times$  & $7.07\times$ \\
3  & 107 & 19 & $5.6\times$ & 48  & 166 & $4.13\times$  & $8.75\times$ \\
5  & 74  & 13 & $5.7\times$ & 33  & 139 & $6.06\times$  & $12.80\times$ \\
10 & 41  & 7  & $5.9\times$ & 19  & 95  & $10.40\times$ & $23.25\times$ \\
20 & 21  & 3  & $7.0\times$ & 7   & 55  & $26.77\times$ & $55.77\times$ \\
\bottomrule
\end{tabular}
\end{table*}

\revzk{Under the simulated separable telemetry, budgets fall $5.5$--$7.0\times$ over tested $k\in\{1,2,3,5,10,20\}$. At $k{=}10$, cost advantages over the full-audit model range from 9.43 to 10.40 for the replay-only model and 22.57 to 23.25 for the integer-checker model across the three attack classes. The traceless control instead spends budget on an uninformative high stratum and can be worse than uniform. Figure~\ref{fig:frontier} and Table~\ref{tab:app_frontier} in Appendix~\ref{app:stratified} show the simulated direct-hit--cost frontier and quantile sensitivity. These characterize conditional operating points, not measured full-system speedups.}

\paragraph{Checkpoint-Interval Pareto.}
We sweep $g\in\{1,2,5,10,20\}$, pricing the online sealed checkpoint-write share
(measured 365\,MB/s write bandwidth) against the expected deduplicated offline
replay at reference budget $s{=}10$; Table~\ref{tab:gpareto} reports the results,
and Appendix~\ref{app:checkpoint_pareto} adds a storage-only sweep over
\revzk{$g\in\{10,50,100,500\}$.} $g{=}5$ keeps the online share below 2\% (0.96\% at 7B,
1.51\% at 14B) \revzk{with about 28 expected replayed steps per audit}; $g\in\{1,2\}$
violates the 2\% online bound, and $g{\ge}10$ nearly doubles the offline burden.
\revzk{This model motivates $g{=}5$ for the BF16 prototype; it is not a full-system timing sweep at each $g$.}

\begin{table}[!htbp]
\centering
\caption{\revzk{\textbf{Checkpoint-interval cost model ($N{=}200$, $s{=}10$).} Online share and replay times use measured bandwidth and step-time inputs. $\E|\bigcup|$ comes from the sampler; parentheses give the uniform-target reference $s(g{+}1)/2$, not a bound for nonuniform targets.}}
\label{tab:gpareto}
\footnotesize
\resizebox{\columnwidth}{!}{%
\begin{tabular}{ccccc}
\toprule
$\mathbf{g}$ & \textbf{Online share 7B / 14B} & $\E|\bigcup|$ & \textbf{Replay 7B} & \textbf{Replay 14B} \\
\midrule
1  & 4.79\% / 7.55\% & 10.0 (10)  & 34.6\,s  & 58.0\,s \\
2  & 2.39\% / 3.77\% & 15.6 (15)  & 53.9\,s  & 90.5\,s \\
\textbf{5}  & \textbf{0.96\% / 1.51\%} & \textbf{27.7 (30)} & \textbf{95.8\,s} & \textbf{160.7\,s} \\
10 & 0.48\% / 0.75\% & 47.2 (55)  & 163.2\,s & 273.8\,s \\
20 & 0.24\% / 0.38\% & 85.7 (105) & 296.5\,s & 497.2\,s \\
\bottomrule
\end{tabular}%
}
\end{table}

\section{Telemetry-Stratified Sampling: Full Grid and Sensitivity}
\label{app:stratified}

\revzk{This appendix extends the simulated-telemetry model in Table~\ref{tab:stratified} to three anomaly-response classes, a traceless control, $N{=}1000$, and quantile sensitivity. These are sampler calculations with injected telemetry, not additional real training runs; the measured rankings are reported in Appendix~\ref{app:real-audit}.} All
cells use $g{=}5$, $\alpha{=}0.2$, $q{=}15\%$, 24 injection seeds at $N{=}200$
(12 at $N{=}1000$); $s^*$ uses the exact two-stratum hypergeometric accounting
of Eq.~\eqref{eq:stratified} (the conservative specification bound differs by at
most 2 steps, e.g., 26 vs.\ 24 at $N{=}200, k{=}2$). Direct-hit probability at $s^*$ is
0.900--0.971 analytically, agreeing with $10^5$-trial process-level Monte-Carlo
within 0.31\% and with the real \texttt{StratifiedBeaconSampler} (10{,}000
\revzk{trials, fresh simulated challenge randomness per trial)} within 0.50\%.

\subsection{\revzk{Per-Class Budgets and Modeled Cost Ratios ($N{=}200$)}}

\begin{table}[!htbp]
\centering
\caption{\revzk{\textbf{Simulated-telemetry budgets and modeled cost ratios by tamper class.}}
Uniform $s^*$ is class-independent (180/137/107/74/41/21 for $k{=}1,2,3,5,10,20$);
traceless $s^*$ is 183/146/121/92/65/48. The replay-only modeled cost ratio is $N/\E|\bigcup|$
(identical across 7B/8B/14B); the integer-checker model prices the 3.32 CPU-s \texttt{nfzk} unit
check. \revzk{The 3.32 CPU-s input is a residual calibration, not standalone checker timing.}}
\label{tab:app_stratified}
\footnotesize
\resizebox{\columnwidth}{!}{%
\begin{tabular}{lcccc}
\toprule
\textbf{Class} & $\mathbf{k}$ & $\mathbf{k_H}$ med & $\mathbf{s^*}$ & \textbf{Replay / Integer} $\times$ \\
\midrule
\texttt{batch\_poison} & 1/2/3/5/10/20 & $=k$ & 33/24/19/13/7/3 & 2.51--26.77 / 5.11--55.77 \\
\texttt{lazy\_skip}    & 1/2/3/5/10/20 & $=k$ & 33/24/19/13/7/3 & 2.52--26.11 / 5.12--55.39 \\
\texttt{lr\_tamper}    & 1/2/3/5/10/20 & $=k$ & 33/24/19/13/7/3 & 2.42--24.23 / 5.06--54.21 \\
\midrule
\texttt{traceless}     & 1/2/3/5/10/20 & 0--3 & 183/146/121/92/65/48 & 1.02--1.91 / 1.08--3.61 \\
\bottomrule
\end{tabular}%
}
\end{table}

\revzk{The simulated anomaly responses place nearly every injected step in the high stratum, yielding the same tabulated budgets. Cost ratios still differ across classes. This controlled separability does not hold for all real deviations.} The traceless control
leaves no informative telemetry signature, so the high stratum does not
reliably concentrate marked steps. The tabulated budgets exceed uniform's
because preferential allocation spends checks on this uninformative stratum.

\subsection{$N{=}1000$ Replication}

At $N{=}1000$ ($|\mathcal{H}|{=}150$, 12 seeds, \texttt{batch\_poison}),
stratified sampling retains its budget advantage over uniform:
$s^* = 168/126/98/67/37/18$ ($s/N \approx 2$--$17\%$) versus uniform
$900/684/536/369/205/108$, giving modeled cost ratios of $2.48$--$19.30\times$ in the replay-only model and
$5.03$--$44.57\times$ in the integer-checker model; the traceless arm requires
$s^* = 915/731/605/463/324/150$, approximately $1.7$--$58.0\%$ higher than uniform.

\subsection{High-Stratum Quantile Sensitivity}

\revzk{The $k{=}1$ budget floor depends on the frozen quantile. An exploratory single-seed probe first injected 20 anomalies and then evaluated a prefix of the labels; its $k{=}5$ case therefore differs from a run with only five injections. This probe gives}
$s^*_{k=1} = 11/22/33/56/109$ and $s^*_{k=5} = 8/8/13/22/42$ at
$|\mathcal{H}| = 10/20/30/50/100$ ($q = 5/10/15/25/50\%$), i.e., the small-$k$
replay-only cost-ratio ceiling scales with the stratum size, not with a sampling-theory limit.

\begin{table}[!htbp]
\centering
\caption{\revzk{\textbf{Modeled cost advantages vs.\ stratum quantile $q$}} ($N{=}200$,
$g{=}5$, $\alpha{=}0.2$, $P_{\mathrm{hit}}{\ge}90\%$, 24 injection seeds per cell).
The minimum observed capture across the 24 seeds is $k_H{=}k$; $s^*$ is class-independent; modeled cost-ratio ranges span
\texttt{batch\_poison}, \texttt{lazy\_skip}, and \texttt{lr\_tamper}. The
pre-registered $q{=}15\%$ arm is Table~\ref{tab:stratified}.}
\label{tab:app_qsens}
\footnotesize
\begin{tabular}{lcccc}
\toprule
$q$ ($|\mathcal{H}|$) & $\mathbf{k}$ & $\mathbf{s^*}$ & \textbf{Replay} & \textbf{Integer} \\
\midrule
5\% (10)  & 1 & 11 & 5.63--5.89$\times$  & 14.07--14.28$\times$ \\
          & 2 & 8  & 7.74--9.08$\times$  & 19.34--20.33$\times$ \\
          & 3 & 6  & 12.11--12.22$\times$ & 27.11--27.18$\times$ \\
          & 5 & 3  & 26.30--33.03$\times$ & 55.50--58.81$\times$ \\
\midrule
7\% (15)  & 1 & 17 & 3.91--4.11$\times$  & 9.31--9.46$\times$ \\
          & 2 & 12 & 5.72--6.19$\times$  & 13.32--13.64$\times$ \\
          & 3 & 9  & 8.18--8.84$\times$  & 18.14--18.54$\times$ \\
          & 5 & 6  & 14.05--15.12$\times$ & 28.25--28.79$\times$ \\
\bottomrule
\end{tabular}
\end{table}

\revzk{We separately run the full 24-seed simulation with the actual requested $k$ injected anomalies at $q\in\{5,7\}\%$. At $q{=}5\%$, $k{=}5$, this gives $s^*{=}3$ rather than the exploratory probe's 8 because high-stratum capture differs (5 versus 2). Floating-point rounding in the implementation makes the nominal $q{=}7\%$ arm use 15 of 200 records. The sweep spans the three simulated anomaly classes,} and Table~\ref{tab:app_qsens} reports the results. Smaller high strata reduce the budget when they retain the marked steps: at $q{=}7\%$, the integer-checker model gives ratios of $9.3$--$28.8\times$ for $k\le5$, and $q{=}5\%$ gives at least $14\times$. In the replay-only model, the $k{=}1$ budget scales approximately with $|\mathcal{H}|$. For several marked steps, the discrete budget can fall sharply once $s_H>|\mathcal{H}|-k_H$, because a high-stratum miss becomes impossible. These gains rely on capture being preserved as the stratum shrinks. The $N{=}1000$ simulation agrees in direction, while the traceless control still requires more samples than uniform. We retain the fixed $q{=}15\%$ configuration for the main experiments.

\subsection{Direct-Hit Probability Frontier ($P^*$ Sweep)}
\label{app:frontier}

\begin{figure*}[t]
\centering
\includegraphics[width=\textwidth]{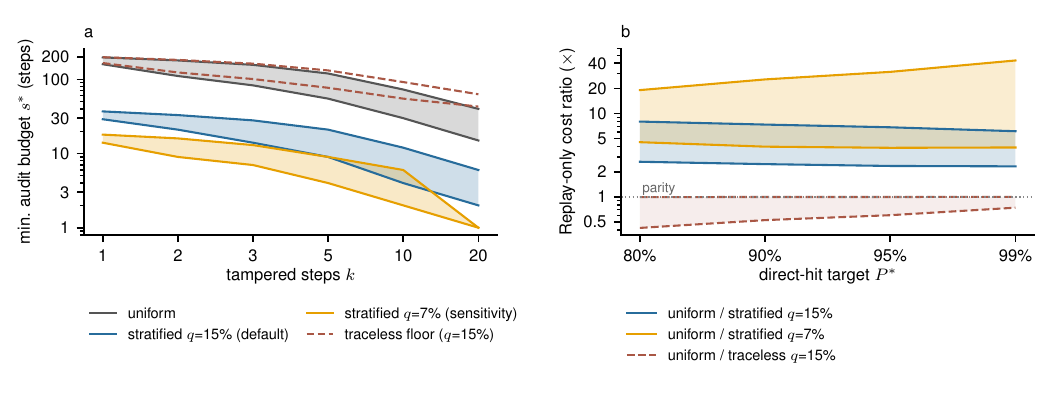}
\caption{\revzk{\textbf{Simulated direct-hit--cost frontier ($N{=}200$, $g{=}5$,}
$\alpha{=}0.2$, \texttt{batch\_poison}, minimum observed $k_H$ over 24 seeds).}
\textbf{a}~Minimum audit budget $s^*$ vs.\ tampered steps $k$; each band
spans $P^*\in\{80,90,95,99\}\%$. The stratified bands sit strictly below
the uniform band in every cell---the frontier shifts down---while the
traceless floor (dashed) rides just above uniform.
\revzk{\textbf{b}~Modeled replay-only cost advantage relative to uniform (uniform cost divided by arm cost)}
(envelope over $k\in\{1,2,3,5,10,20\}$): the stratified arms clear parity by
$2.3$--$43.2\times$ everywhere, and the traceless floor stays below it
(worst $0.42\times$), converging to $\approx1.0\times$ at
$P^*{=}99\%,k{\le}3$.}
\Description{Two-panel plot. Panel a shows log-scale minimum audit
budget versus tampered steps for uniform, stratified q=15\%, stratified
q=7\% arms as bands over direct-hit targets 80--99\%, plus a dashed
traceless floor slightly above uniform. Panel b shows the replay-only model cost
ratio to uniform versus direct-hit target for the same arms, with the
traceless band below the parity line.}
\label{fig:frontier}
\end{figure*}

We sweep the direct-hit target $P^*\in\{80,90,95,99\}\%$ on the protocol of
Table~\ref{tab:stratified} ($N{=}200$, $g{=}5$, $\alpha{=}0.2$, minimum observed $k_H$
over 24 injection seeds, \texttt{batch\_poison} as the representative class);
Figure~\ref{fig:frontier} and Table~\ref{tab:app_frontier} present the results.
\texttt{lazy\_skip} and \texttt{lr\_tamper} share \texttt{batch\_poison}'s $s^*$ in
all but two $(q,k)$ combinations. At $q{=}15\%$, $k{=}20$, $P^*\in\{95,99\}\%$,
\texttt{batch\_poison} needs 4/6 samples where the other classes need 3/4; at
$q{=}7\%$, $k{=}10$, $P^*\in\{90,95,99\}\%$, it needs 3/4/6 where they need 2/3/4.
The plotted frontier is therefore conservative. In every $(P^*,k)$ cell of both stratified arms the budget
$s^*$ sits strictly below uniform (panel a: three separated bands); the
$q{=}7\%$ frontier lies below the $q{=}15\%$ frontier throughout
(typically $\approx2\times$ lower $s^*$). The edge steepens at
$P^*{=}99\%$: once
$s_H{\ge}|\mathcal{H}|{-}k_H{+}1$ a miss inside
$\mathcal{H}$ is impossible, so stratified $s^*$ nearly stops growing
($q{=}7\%$, $k{=}1$: 18 at both 95\% and 99\%; $k{=}20$: $s^*{=}1$ at
every target) while uniform keeps climbing toward $N$. The traceless
floor marks the boundary of the mechanism: its $s^*$ exceeds uniform
in every cell (replay-only model ratio at or below parity throughout, worst
$0.42\times$ at $q{=}15\%$, $k{=}20$, $P^*{=}80\%$), because the
$(1{-}\alpha)$ high-stratum budget samples an anomaly-free layer. The
penalty narrows as $P^*{\to}99\%$, reaching $0.99$--$1.00\times$ parity
at $k{\le}3$ with a $0.74$--$0.88\times$ residual at $k{\ge}10$. The $q{=}7\%$ traceless floor is never worse than
$q{=}15\%$ (residual $0.84$--$1.00\times$ at $P^*{=}99\%$). Analytic $s^*$ across all 192
frontier cells agrees with $10^5$-trial process-level Monte-Carlo within
$0.40\%$ and with the real \texttt{StratifiedBeaconSampler} (10{,}000
trials on boundary cells) within $0.06\%$.

\begin{table}[!htbp]
\centering
\caption{\textbf{Minimum audit budget $s^*$ across the direct-hit
frontier} ($N{=}200$, $g{=}5$, $\alpha{=}0.2$, \texttt{batch\_poison},
minimum observed $k_H$ over 24 seeds). Uniform $s^*$ is class-independent;
\texttt{lazy\_skip} and \texttt{lr\_tamper} differ in two $(q,k)$ combinations (see
text). The traceless floor is the degradation boundary: its $s^*$
exceeds uniform everywhere, yet its replay-only cost ratio approaches parity
as $P^*{\to}99\%$ for $k{\le}3$ ($0.74$--$0.88\times$ residual at
$k{\ge}10$).}
\label{tab:app_frontier}
\footnotesize
\resizebox{\columnwidth}{!}{%
\begin{tabular}{llcccccc}
\toprule
\textbf{Arm} & $P^*$ & $k{=}1$ & 2 & 3 & 5 & 10 & 20 \\
\midrule
uniform & 80\% & 160 & 111 & 83 & 55 & 30 & 15 \\
        & 90\% & 180 & 137 & 107 & 74 & 41 & 21 \\
        & 95\% & 190 & 155 & 126 & 90 & 51 & 27 \\
        & 99\% & 198 & 180 & 157 & 120 & 73 & 40 \\
\midrule
stratified $q{=}15\%$ & 80\% & 29 & 21 & 14 & 9 & 4 & 2 \\
        & 90\% & 33 & 24 & 19 & 13 & 7 & 3 \\
        & 95\% & 36 & 28 & 23 & 16 & 8 & 4 \\
        & 99\% & 37 & 33 & 28 & 21 & 12 & 6 \\
\midrule
stratified $q{=}7\%$ & 80\% & 14 & 9 & 7 & 4 & 2 & 1 \\
        & 90\% & 17 & 12 & 9 & 6 & 3 & 1 \\
        & 95\% & 18 & 14 & 11 & 7 & 4 & 1 \\
        & 99\% & 18 & 16 & 13 & 9 & 6 & 1 \\
\midrule
traceless $q{=}15\%$ & 80\% & 166 & 124 & 101 & 77 & 55 & 43 \\
        & 90\% & 183 & 146 & 121 & 92 & 65 & 48 \\
        & 95\% & 192 & 162 & 137 & 106 & 73 & 53 \\
        & 99\% & 199 & 183 & 163 & 132 & 92 & 63 \\
\bottomrule
\end{tabular}%
}
\end{table}

\section{Simulated Injection Grid and Statistical Calibration}
\label{app:injection_grid}

\subsection{Experimental Parameter Space}
The experimental grid evaluates seven injection families (six attacks covering
A1--A4, A6, and A7, plus a negative control; RNG state $\rho_i$ is included
in the bound checkpoint, but A5 is not injected) across
\revzk{the following dimensions:}
\begin{itemize}[leftmargin=*]
\item \textbf{Attack Families (7):} \textsf{skip\_step} (A1), \textsf{swap\_base} (A2),
\textsf{alter\_hparams} (A6), \textsf{single\_step\_poison} (A4), \textsf{forge\_checkpoint} (A3/A7),
\textsf{uncommitted\_data} (A6), and \textsf{committed\_poison} (Negative Control).
\item \textbf{Binding Configurations (2):} Full binding vs.\ ablated one-binding-off
(testing failures exposed by the selected binding removals).
\item \textbf{Perturbation Intensity ($k_{\mathrm{inj}}$):} $k_{\mathrm{inj}} \in \{1, 5, 10\}$ injected violating units.
\item \textbf{Sampling Strategies (10):} Uniform-fixed-$s$ ($s \in \{2,10,20,40\}$), Bernoulli sampling ($p_{\mathrm{inc}} \in \{0.01,0.05,0.10,0.20\}$), block-stratified sampling, and input-aware probability-proportional-to-size sampling without replacement (PPS; $s{=}10$).
\item \textbf{Random Seeds and Repetitions:} Seeds $\{42, 1024, 2048\}$ evaluated over
500 trials per cell (630{,}000 total simulated audit trials).
\end{itemize}

\subsection{Wilson Interval Coverage Analysis}
For $x$ detections in $n$ trials, let $\hat p=x/n$. We use the 95\% Wilson score interval~\cite{wilson1927},
\begin{equation}
 \frac{\hat p+\frac{z^2}{2n}\;\pm\;z\sqrt{\tfrac{\hat p(1-\hat p)}{n}+\tfrac{z^2}{4n^2}}}{1+z^2/n},\quad z=1.96.
 \label{eq:wilson}
\end{equation}
For each configuration, we compare $\hat{p}=x/500$ with a reference probability. The 1{,}134 exact references use Eq.~\eqref{eq:hypgeom} for uniform-fixed-$s$, Eq.~\eqref{eq:bernoulli} for Bernoulli sampling, or the independent-inclusion formula for the block-stratified arm:
\begin{equation}
 P_{\mathrm{detect}}=1-\prod_{i\in V}(1-p_i),
 \label{eq:block-detection}
\end{equation}
where $p_i$ is the registered inclusion probability of step $i$ (0.02/0.05/0.08/0.05 in the four 50-step blocks). The remaining 126 input-aware PPS references are estimates from 2{,}000 Monte Carlo trials per cell. The empirical Wilson intervals cover 1{,}118/1{,}134 exact references and 125/126 Monte Carlo estimates, totaling $1{,}243/1{,}260=98.65\%$. For the latter cells, this comparison does not account for uncertainty in the reference estimate. Table~\ref{tab:app_outliers}
shows five calibration examples: the first two theoretical probabilities fall
outside their Wilson intervals, and the other three are covered, so the table
illustrates boundary behavior rather than listing outliers. Coverage alone does
not establish absence of bias.

\begin{table}[!htbp]
\centering
\caption{\textbf{Calibration examples from the 1{,}260-cell grid.}
The first two theory values lie outside the displayed intervals; the
remaining three lie inside.}
\label{tab:app_outliers}
\resizebox{\columnwidth}{!}{%
\begin{tabular}{lccccc}
\toprule
\textbf{Attack Arm} & $\mathbf{k_{\mathrm{inj}}}$ & \textbf{Strategy} & \textbf{Theory} $\mathbf{\Pdetect}$ & \textbf{Empirical} $\mathbf{\hat{p}}$ & \textbf{95\% Wilson CI} \\
\midrule
\texttt{skip\_step} & 1 & $s{=}10$ (Seed 42) & \revzk{0.0977} & 0.0700 & \revzk{$[0.0508, 0.0958]$} \\
\texttt{single\_step\_poison} & 1 & $s{=}10$ (Seed 42) & \revzk{0.0977} & 0.0700 & \revzk{$[0.0508, 0.0958]$} \\
\texttt{uncommitted\_data} & 1 & $s{=}2$ (Seed 1024) & 0.0100 & 0.0140 & $[0.0068, 0.0286]$ \\
\texttt{uncommitted\_data} & 5 & $p_{\mathrm{inc}}{=}0.05$ (Seed 2048) & 0.2262 & 0.2100 & $[0.1766, 0.2478]$ \\
\texttt{alter\_hparams} & 1 & $s{=}2$ (Seed 42) & 0.0199 & 0.0120 & $[0.0055, 0.0259]$ \\
\bottomrule
\end{tabular}%
}
\end{table}

\subsection{Calibrated Detection Slice}
\label{app:e5_slice}

\begin{table*}[t]
\centering
\caption{\textbf{Adversarial attack detection at $N{=}200, s{=}10$ across 500 trials per seed (seeds 42/1024/2048).} $|V|$ denotes violation sets assigned by the sampling simulator. Cascading sets require an interval-check predicate and do not establish the coverage of a local-step verifier. $^\dagger$Denotes borderline Wilson CI boundary cells.}
\label{tab:e5}
\resizebox{\textwidth}{!}{%
\begin{tabular}{llcccl}
\toprule
\textbf{Attack Family} & $\mathbf{k_{\mathrm{inj}}}$ & $\mathbf{|V|}$ \textbf{(3 Seeds)} & \textbf{Theoretical} $\mathbf{\Pdetect}$ & \textbf{Empirical Detection} & \textbf{Rejecting Mechanism} \\
\midrule
\texttt{skip\_step} (A1) & 1 & 2 / 47 / 39 & 9.8 / 93.6 / 89.2\% & 7.0$^\dagger$ / 93.0 / 90.6\% & Chain adjacency link failure ($f7$) \\
\texttt{skip\_step} (A1) & 5 & 67 / 91 / 77 & 98.5 / 99.8 / 99.3\% & 99.4 / 99.8 / 99.4\% & Step index mismatch ($f3$) \\
\texttt{skip\_step} (A1) & 10 & 141 / 138 / 154 & $\approx 100\%$ & 100\% & Intermediate chain discontinuity \\
\texttt{swap\_base} (A2) / \texttt{alter\_hparams} (A6) & 5 & 67 / 91 / 77 & 98.5 / 99.8 / 99.3\% & 99.4 / 99.8 / 99.4\% & Spec fingerprint mismatch ($\sigid, C_{\mathrm{base}}, \mathrm{hp}$) \\
\texttt{single\_step\_poison} (A4) & 1 & 2 / 47 / 39 & 9.8 / 93.6 / 89.2\% & 7.0$^\dagger$ / 93.0 / 90.6\% & Output boundary mismatch ($\bind_{\mathrm{out}}$) \\
\texttt{forge\_checkpoint} (A3/A7) & 1 & 50 / 51 / 1 & 94.8 / 95.1 / 5.0\% & 94.6 / 94.0 / 5.6\% & State anchor rejection (\textsf{CHECKPOINT\_MISMATCH}) \\
\texttt{forge\_checkpoint} (A3/A7) & 5 & 200 / 200 / 200 & 100\% & 100\% & State anchor rejection (\textsf{CHECKPOINT\_MISMATCH}) \\
\texttt{uncommitted\_data} (A6) & 1 / 5 / 10 & $|V|{=}k$ & 5.00 / 22.83 / 40.85\% & 3.2--5.2 / 21.8--22.0 / 39.6--42.2\% & Dataset membership proof fails against $R_D$ \\
\midrule
\texttt{committed\_poison} (Neg.\ Ctrl) & any & 0 & 0\% & 0\% & Outside the simulated violation predicate \\
\bottomrule
\end{tabular}%
}
\end{table*}

\subsection{Conditional Economic Deterrence}
\label{app:deterrence}
In a risk-neutral, single-deviation model, let $G$ be the gain received only upon nondetection and $\Lambda$ the enforceable loss upon detection. The expected payoff is negative when
\begin{equation}
 \begin{split}
 (1-\Pdetect)G-\Pdetect\Lambda&<0\\
 \Longleftrightarrow\quad \Lambda/G&>(1-\Pdetect)/\Pdetect.
 \end{split}
 \label{eq:deterrence}
\end{equation}
At $N{=}200$ and $s{=}10$, the uniform probabilities for $|V|{=}1,5,10$ require ratios above 19, 3.38, and 1.45, respectively. This model assumes enforceable settlement and excludes repeated-game and risk-preference effects. The experiments produce rejection evidence but do not deploy escrow or measure deterrence.

\section{\revzk{Auxiliary Cross-Architecture Algebraic Checks}}
\label{sec:eval-freivalds}

\revzk{The auxiliary \texttt{nfzk} checker evaluates exact-integer relations across CPU and GPU hardware. Our tested BF16 GPU/CPU paths, by comparison, use different reduction orders: cuBLAS and oneDNN results mismatch on} $0.042$--$0.195\%$ of GEMM
elements at $d\in\{1024,4096\}$ (an existence result, not a convergence claim).
The integer-domain witness carries no ordering ambiguity: at the 7B layer
dimension $d{=}4096$ (LoRA $r{=}16$), a GPU-generated exact-integer product
witness is bitwise equal to CPU recomputation ($0/16{,}777{,}216$ mismatches),
and batched Freivalds verification costs 50\,ms against 47.9\,s of
exact-integer $O(d^3)$ recomputation, a $951\times$ reduction ($3.7\times$
against the fp64-exact path). A single-element claim tamper is caught
in $12/12$ trials at $R{=}8$ rounds, and an AdamW-chain check over the
$2rd$-element LoRA surface costs 331\,ms per projection, which extrapolates to
\revzk{about 65\,s over 28 layers and seven projections per layer at $r{=}16$. This is a shape-based extrapolation, not a full-step measurement of the main $r{=}8$, q/k/v/o training workload. Table~\ref{tab:freivalds} summarizes operator timings.}

\begin{table}[!htbp]
\centering
\caption{\textbf{Auxiliary \texttt{nfzk} Freivalds microbenchmark} (CPU, min-of-reps). $O(d^3)$ is exact-integer recomputation; FV is batched $R{=}8$-round verification of the opened integer product claim, evaluated in fp64. \revzk{Detection uses int64 and independent binary challenges; the 99/100 outcomes include one all-zero challenge row that misses the changed element. The fp64 timings are a separate path.} $^\dagger$fp64-FV versus fp64-exact recomputation; the $d{=}64$ fp64 path is a constant-overhead regime (excluded).}
\label{tab:freivalds}
\footnotesize
\begin{tabular}{ccccc}
\toprule
$\mathbf{d}$ & $\mathbf{O(d^3)}$ \textbf{i64} & \textbf{FV fp64} & \textbf{Speedup vs.\ i64} & \textbf{Detect} \\
\midrule
64   & 0.2\,ms      & 0.07\,ms & $2.8\times$   & 99/100 \\
1024 & 703\,ms      & 40\,ms   & $17.6\times$  & 99/100 \\
2048 & 5{,}957\,ms  & 46\,ms   & $129\times$   & 30/30 \\
4096 & 47{,}910\,ms & 50\,ms   & $951\times$ ($3.7\times^\dagger$) & 12/12 \\
\bottomrule
\end{tabular}
\end{table}

\section{\revzk{Component-Priced Resource Ledger and Parallel Scaling}}
\label{app:ledger}

\begin{table*}[t]
\centering\color{black}
\caption{\textbf{Modeled audit resource ledger across sampling rate $s/N$ at $N{=}200, g{=}50$.} Proving is priced at the measured 273.33 GPU-s/unit and recomputation at 1.70 GPU-s/step. Modeled cost ratios compare different verification goals (probabilistic and $|V|$-dependent for sampled rows, deterministic for full rows). $^\dagger$Assumed 1.2 CPU-s/unit binding-level check, not a measurement. The tabulated totals exclude verifier-side interval replay of $\Sigma\mathrm{dist}$ steps (0.25 CPU-s/step on the tiny-$\sigma$ fixture).}
\label{tab:e4}
\resizebox{\textwidth}{!}{%
\begin{tabular}{lcccccccc}
\toprule
\textbf{Sampling Rate} & $\mathbf{s}$ & $\mathbf{\Sigma\mathrm{dist}}$ & \textbf{Proving model (GPU-s)} & \textbf{Replay model (GPU-s)} & \textbf{GPU total model (GPU-s)} & \textbf{CPU check model (CPU-s)}$^\dagger$ & \textbf{Size estimate} & \textbf{Full / arm cost} \\
\midrule
1\% & 2 & 16 & 546.7 & 27.2 & 573.9 & 2.4 & 2.86\,MB & $110.4\times$ \\
\textbf{5\% (Calibrated)} & \textbf{10} & \textbf{205} & \textbf{2{,}733.3} & \textbf{348.5} & \textbf{3{,}081.8} & \textbf{12.0} & \textbf{14.29\,MB} & $\mathbf{20.55\times}$ \\
10\% & 20 & 421 & 5{,}466.6 & 715.7 & 6{,}182.3 & 24.0 & 28.58\,MB & $10.24\times$ \\
20\% & 40 & 915 & 10{,}933.2 & 1{,}555.5 & 12{,}488.7 & 48.0 & 57.16\,MB & $5.07\times$ \\
\midrule
Full-coverage model ($\mathrm{B}_{\mathrm{Full}}$) & 200 & 5{,}100 & 54{,}666.0 & 8{,}670.0 & 63{,}336.0 & 240.0 & 285.81\,MB & $1.00\times$ \\
Per-step model ($\mathrm{B}_{\mathrm{Step}}$) & 200 & -- & 54{,}666.0 & -- & 54{,}666.0 & -- & -- & $1.16\times$ \\
\bottomrule
\end{tabular}%
}
\end{table*}

\revzk{This appendix reports a component-priced operating-point model. Its 273.33\,GPU-s/unit proving input comes from an external small-model pilot that completed 15 of 17 components; two attention components failed and the coverage matrix records further semantic gaps. It does not price a completed private step relation. The 1.2\,CPU-s binding cost is an assumption. Cross-goal ratios compare different detection guarantees and must not be used as same-goal experimental speedups.}

{\color{black}
\paragraph{100-Step Training Measurement and Audit Cost Model.} We combine a measured
100-step 1.5B run with a protocol-priced audit ledger. The baseline finishes in
87.53\,s and train-and-commit in 89.03\,s ($+1.73\%$; boundary records $+1.15\%$,
commit calls $+0.0016\%$). Both arms share kernels, so the determinism premium of
Section~\ref{sec:eval-commit} is excluded. After sealing, the ledger prices the
audit samples $Q = \{6, 11, 28, 39, 100\}$ derived from a fixed all-zero development beacon value and the sealed chain root at
1{,}483.8\,GPU-s of modeled proving and recomputation, giving a full-audit/sample cost ratio of $19.92\times$ against the 29{,}562.6\,GPU-s ledger. Under a serial accounting assumption, adding training wall time and $5\times1.2$ CPU-s of binding checks gives a 1{,}578.8\,s time proxy. Applying the per-step component price and CPU-check allowance to every step gives 27{,}540.5\,s, a ratio of $17.44\times$. These proxies add GPU work, CPU work, and training time without overlap; they are not measured elapsed times. The proving input covers an incomplete relation, and the ratios compare different verification goals. Section~\ref{sec:eval-measured} reports measured audits at matched direct-hit targets.

\paragraph{Calibrated Ledger Evaluation.} At the pre-registered benchmark
($N{=}200, g{=}50, s/N{=}5\%$, i.e., $s{=}10$), Table~\ref{tab:e4} gives the full
ledger: the model assigns 3{,}081.8\,GPU-s (2{,}733.3 proving plus 348.5 recomputation
across $\Sigma\mathrm{dist}{=}205$ re-executed steps) and 12.0\,CPU-s verification,
versus 63{,}336.0\,GPU-s for the full-audit ledger ($\mathrm{B}_{\mathrm{Full}}$, $20.55\times$)
and 54{,}666.0\,GPU-s for direct per-step proving
($\mathrm{B}_{\mathrm{Step}}$, $17.74\times$). These cross-goal ratios are
dominated by the sampling fraction ($\approx N/s$): the sampling calculation gives the goal
$(k_{\min}{=}1, P^*{=}5.00\%)$ under Eq.~\eqref{eq:hypgeom} with the enforced
interval predicate, whereas both baseline ledgers price checks of every step under their component relation. Booking verifier-side interval replay re-prices the 5\% ratios to
$18.46\times$ against the original full-audit ledger, or $20.99\times$ when that baseline also includes replay; these are different denominators, not a statistical interval. The ratio is $15.94\times$ against $\mathrm{B}_{\mathrm{Step}}$; the full-coverage comparator explicitly sets $s{=}N$ and costs more than direct per-step proving under this ledger. Full coverage differs from merely guaranteeing one hit: for a known marked count $k\ge1$, uniform sampling reaches hit probability one at $s{=}N-k+1$. The worst case $k{=}1$ requires $N$ samples. The component-derived storage estimate falls from 285.8\,MB to 14.3\,MB at 5\% sampling; it is not a full-size private-proof measurement.

\paragraph{Parallel Audit Scaling.}
\label{sec:eval-parallel}
Because audit units carry estimated work weights $w_i = \mathrm{dist}(i)\cdot t_{\mathrm{step}} + t_{\mathrm{prove}}$,
the $s$ jobs can run independently across provider GPUs. The measured workload restores intervals, constructs opened witnesses, and checks the integer relation; the code retains its historical prove/verify API names.
On a 4-GPU A100-40GB host with the two-layer model ($s{=}10$, $\Sigma\mathrm{dist}{=}205$),
median wall-clock falls from 1{,}040.1\,s (1 GPU) to 576.2\,s (2 GPUs, $1.80\times$)
and 313.2\,s (4 GPUs, $\mathbf{3.32\times}$, 83.0\% efficiency), and all 120 audited
runs verify bit-exactly. LPT binning shrinks straggler tails from 62.2\,s to 30.0\,s,
bringing this two-layer audit workload to about 5 minutes on four GPUs.

}

\section{\revzk{Requirements for a Future Private Audit Backend}}
\label{app:circuits}

\revzk{The current audit discloses the states and batches required for replay. A private backend would replace those openings with a proof of the same computation and bindings. The following obligations describe this extension; neither the opened-witness checker nor the external component pilot implements a complete private relation.}

\subsection{\revzk{Computation, Binding, and Privacy Obligations}}
\revzk{A private audit-unit proof $\pi_i$ would need to establish:}
\begin{enumerate}[leftmargin=*]
\item \revzk{\textbf{The required interval computation.} The relation must cover forward and backward operations, the specified LoRA update, and complete optimizer semantics, including AdamW bias correction, weight decay, scheduling, and RNG dependencies. Numerical encoding and rounding must match the committed specification. Proving only a selected local transition leaves its predecessor unverified; the required interval and any predecessor-telemetry computation must also be covered.}
\item \revzk{\textbf{Sealed state and record bindings.} The proof must bind its starting checkpoint, intervening records, and final state to the acknowledged seal, including the specification fingerprint and telemetry values used by the audit. A proof of unrelated arithmetic does not establish these links.}
\item \revzk{\textbf{Data membership and disclosure.} The relation must establish the prescribed batch positions and membership against $R_D$. A complete privacy argument must state which commitments, telemetry, selections, and outcomes remain public and establish zero knowledge for the remaining witnesses. Hash preimage resistance alone does not provide this guarantee.}
\end{enumerate}
\revzk{Potential backend techniques include Spartan~\cite{spartan}, zkLLM~\cite{zkllm}, and polynomial-commitment systems~\cite{basefold,brakedown,batchzk}. Constraint counts, proving cost, and privacy composition require the complete implemented relation; hash input length and integer-checker microbenchmarks do not determine them.}

\section{Auditor Leakage Surface}
\label{app:privacy}

\revzk{Table~\ref{tab:privacy} summarizes the auditor's view. The implemented paths disclose replay inputs; hiding them requires the additional proof obligations in Appendix~\ref{app:circuits}.}

\begin{table}[!t]
\centering
\caption{\textbf{Auditor leakage surface.} The current opened-witness
audit discloses replay inputs; the \emph{target} row describes a future private-proof backend satisfying Appendix~\ref{app:circuits}.}
\label{tab:privacy}
\footnotesize
\begin{tabularx}{\columnwidth}{@{}>{\raggedright\arraybackslash}p{.27\columnwidth}X@{}}
\toprule
\textbf{Artifact} & \textbf{Auditor View} \\
\midrule
$\sigma,\sigid,$ $R_D,C_{\mathrm{base}}$ & \revzk{Public specification and digests, without a hiding guarantee} \\
Commitment $c_i$ & \revzk{Hiding needs fresh secret uniform salts and a random-oracle argument. BF16 uses public development salts and claims no hiding} \\
Batch Witness $\bind_{x_i}$ & \revzk{Batches are opened. The reference protocol checks Merkle membership; BF16 checks hashes and positions} \\
Reported Telemetry & \revzk{All reported losses, update norms, and available position features used for ranking; fixed by sealing and checked in replayed intervals} \\
Replay Inputs & \revzk{Both checkers receive required checkpoints and interval states} \\
Private proof $\pi_i$ (\emph{target}) & \revzk{Needs a complete zero-knowledge relation and public-input/leakage specification; no end-to-end backend measured} \\
Challenge Set $Q$ & \revzk{Reference: public beacon; BF16: local randomness. Selection may reveal rankings} \\
Delivered Model $\mathbf{W}_{\mathrm{final}}$ & Bound by $h(\mathbf{W}_{\mathrm{final}})$. Replay can open unselected steps, including predecessor intervals \\
\bottomrule
\end{tabularx}
\end{table}

\section{Artifact Availability and Reproducibility Checklist}
\label{app:reproducibility}

\revzk{The retained evidence package contains implementation snapshots, configurations, commitment records, challenges, verdicts, logs, and analysis outputs. The supplementary evaluation archive includes 1{,}689 source files; 172 large checkpoint/state tensor paths are listed with their source locations and digests but are not bundled. Consequently, the package supports inspection of recorded decisions and recomputation of reported summaries, not independent execution of every opened-witness audit.}

\revzk{The record-checking entry point is \path{check_experiments_20261030.py} in the artifact directory. It checks stored evidence and recomputes the associated calculations. Re-executing the BF16 audits additionally requires the omitted tensors, matching model/data revisions, and a compatible GPU/software environment. A complete anonymous executable package still requires assembly and verification.}

\section{Ethical Considerations}
\label{app:ethics}

This paper investigates independent auditing of outsourced model adaptation.
All empirical evaluations were performed using publicly accessible, open-license datasets
(Databricks Dolly-15k~\cite{dolly}) and open-weights models (Qwen2.5-1.5B, Qwen2.5-7B, Qwen2.5-14B, and Llama-3.1-8B-Instruct).
\revzk{Attacks were injected into controlled training and audit runs using public data. Another service remained resident on the GPU during the 7B prototype run; its latency was not measured, so service continuity does not establish absence of performance impact. No private client data or human-subject experiments were used.} The proposed auditing framework
promotes algorithmic accountability and verifiable execution in outsourced machine learning
services. The current audit discloses the checkpoints, states, and batches needed for
interval replay; avoiding disclosure of proprietary weights or private client data is the goal of
the target private-proof interface.

\end{document}